\documentclass[a4paper]{jpconf}
\usepackage{graphicx}
\usepackage{bm}

\begin{document}

\hspace*{12cm} 

\newcommand{\ltorder}
      {\ifmmode       { \raisebox{-.4em}{$<$}\atop\sim}
         \else        {$\raisebox{-.4em}{$<$}\atop\sim$}
      \fi}
\newcommand{\gtorder}
      {\ifmmode       { \raisebox{-.4em}{$>$}\atop\sim}
         \else        {$\raisebox{-.4em}{$>$}\atop\sim$}
      \fi}

\title{Unpolarized Structure Functions at Jefferson Lab}

\author{M.~E.~Christy$^1$ and W.~Melnitchouk$^2$}
\address{
$^1$ Hampton University, Hampton, Virginia 23668, USA \\
$^2$ Jefferson Lab, Newport News, Virginia 23606, USA}

\begin{abstract}
Over the past decade measurements of unpolarized structure functions with
unprecedented precision have significantly advanced our knowledge of
nucleon structure.  These have for the first time allowed quantitative
tests of the phenomenon of quark-hadron duality, and provided a deeper
understanding of the transition from hadron to quark degrees of freedom
in inclusive scattering.  Dedicated Rosenbluth-separation experiments
have yielded high-precision transverse and longitudinal structure
functions in regions previously unexplored, and new techniques have
enabled the first glimpses of the structure of the free neutron, without
contamination from nuclear effects.
\end{abstract}

\vspace*{-0.5cm}
  
\section{Introduction}
\label{sec:intro}

Throughout the modern history of nuclear and particle physics,
measurements of structure functions in high-energy lepton-nucleon
scattering have played a pivotal role.  The demonstration of structure
function scaling in early deep-inelastic scattering (DIS) experiments
at SLAC in the late 1960's \cite{DIS} established the reality of quarks
as elementary constituents of protons and neutrons --- a feat recognized
by the award of the 1990 Nobel Prize in Physics to Friedman, Kendall and
Taylor.  This paved the way to the development of quantum chromodynamics
(QCD) as the theory of strong nuclear interactions, and its subsequent
confirmation several years later through the discovery of logarithmic
scaling violations in structure functions \cite{Violation}.

The interpretation of structure functions in terms of quark and gluon
(or parton) momentum distributions resulted in the emergence of a
remarkably simple and intuitive picture of the nucleon \cite{Feynman},
allowing a vast amount of scattering data to be described in terms of a
few universal functions --- the parton distribution functions (PDFs).
At leading order (LO) in $\alpha_s$, the $F_2$ structure function
of the proton, for example, could be simply represented as a charge
squared-weighted sum of PDFs,
\begin{equation}
F_2 = \sum_q e_q^2\, x (q + \bar q)
    = {4\over 9} x (u+\bar u)
    + {1\over 9} x (d+\bar d)
    + {1\over 9} x (s+\bar s) + \cdots\, ,
\end{equation}
where $q$ and $\bar q$ are the quark and antiquark momentum distribution
functions, usually expressed as functions of the momentum fraction $x$
of the nucleon carried by the parton, at a scale given by the momentum
transfer squared $Q^2$.

Over the ensuing decades concerted experimental DIS programs at SLAC,
CERN, DESY and Fermilab have provided a detailed mapping of the PDFs
over a large range of kinematics, with $Q^2$ and $x$ spanning several
orders of magnitude.  To manage the ever increasing number of data sets,
from not only inclusive DIS but also other high energy processes such
as Drell-Yan, $W$-boson and jet production in hadronic collisions at
Fermilab, sophisticated global fitting efforts were developed
\cite{CTEQ,MSTW,Alekhin,NNPDF} that now include perturbative corrections
calculated to next-to-leading order (or higher) in the strong coupling
constant $\alpha_s$.
With the increasing energies available at facilities such as the DESY
$ep$ collider HERA (and in future the Large Hadron Collider at CERN),
PDF studies turned their attention primarily to exploring the region
of very small $x$ (down to $x \sim 10^{-6}$), where the structure of
the nucleon is dominated by its sea quark and gluon distributions.
However, while opening the door to exploration of phenomena such as
saturation and $Q^2$ evolution in new kinematic regimes, one can argue
whether DIS at low $x$ measures the intrinsic structure of the nucleon
or the hadronic structure of the virtual photon, $\gamma^*$.
Because the virtual photon in the DIS process can fluctuate into
$q\bar q$ pairs whose coherence length $\lambda \sim 1/M x$ becomes
large at small $x$, DIS at $x \ltorder 0.1$ really probes the
$\gamma^* N$ interaction rather than the structure of the nucleon
or the $\gamma^*$ separately.  At high $x$, in contrast, the virtual
photon is point-like and unambiguously probes the structure of the
nucleon \cite{Levy}.

Moreover, despite the impressive achievements over the past 4 decades,
there are still some regions of kinematics where our knowledge of
structure functions and PDFs remains unacceptably poor, with little
progress made since the 1970's.  A striking example is the region of
large $x$ ($x \gtorder 0.5$), where most of the momentum is carried
by valence quarks, and sea quarks and gluons are suppressed.
Here the valence quark PDFs can be more directly related to quark models
of hadron structure; however, the rapidly falling cross sections have
made precision measurements extremely challenging.  Another example is
the pre-asymptotic region dominated by nucleon resonances, where data
on the individual transverse and longitudinal cross sections at
intermediate and high values of $Q^2$ are either nonexistent or have
large uncertainties.  With the availability of continuous, high
luminosity electron beams at the CEBAF accelerator, the first decade
of experiments at Jefferson Lab has seen a wealth of high-quality data
on unpolarized structure functions of the nucleon, penetrating into the
relatively unexplored large-$x$ domain and the transition region between
resonances and scaling.

The new data in the resonance region confirmed in spectacular fashion
the phenomenon of quark-hadron duality in the proton $F_2$ structure
function, and revealed intriguing details about the workings of duality
in a number of other observables.  The impact of this has been a
re-evaluation of the applicability of perturbative QCD to structure
functions at low $Q^2$, and has allowed a much larger body of data
to be used in global PDF analyses \cite{CTEQ6X}.  Jefferson Lab has
also set a new standard in the determination of Rosenbluth-separated
longitudinal and transverse structure functions, which eliminates the
need for model-dependent assumptions that have plagued previous
extractions of structure functions from cross section data.

On the theoretical front, the region of large $x$ and low $Q^2$ brings
to the fore a number of issues which complicate structure function
analysis, such as $1/Q^2$ suppressed target mass and higher twist
corrections, and nuclear corrections when scattering from nuclear
targets.  Controlling these corrections requires more sophisticated
theoretical tools to be developed, and has motivated theoretical
studies, many of which are still ongoing.  It has also paved the
way towards the 12~GeV experimental program, in which structure
functions will be measured to very high $x$ in the DIS region,
addressing some long-standing questions about the behavior of PDFs
as $x \to 1$.

In the next section we summarize the kinematics and formalism relevant
for inclusive lepton--nucleon scattering, including the key results
from the operator product expansion.
Measurements of the proton $F_2^p$ structure functions and their
moments are reviewed in Sec.~\ref{sec:F2p}, together with their role
in the verification of quark-hadron duality.
Data on the deuteron $F_2^d$ structure function are presented in
Sec.~\ref{sec:F2d}, and the extraction from these of the neutron $F_2^n$
structure function $F_2^n$ is discussed in Sec.~\ref{sec:F2n}.
Section~\ref{sec:FL} reviews new measurements of the longitudinal
structure function $F_L$, while Sec.~\ref{sec:other} surveys results
from semi-inclusive pion production.
Finally, in Sec.~\ref{sec:outlook} we describe the impact that the
Jefferson Lab data have had on our understanding of nucleon structure in
a global context, and briefly outline prospects for future measurements
in the 12~GeV era.

\section{Formalism}
\label{sec:form}

\subsection{Kinematics}
\label{ssec:kin}

Because of the small value of the electromagnetic fine structure
constant, $\alpha = e^2/4\pi$, the inclusive scattering of an
electron from a nucleon, $e(k) + N(p) \to e'(k') + X$, can usually
be approximated by the exchange of a single virtual photon,
$\gamma^* (q)$, where $q = k'-k$.
In terms of the laboratory frame incident electron energy $E$, the
scattered electron energy $E'$, and the scattering angle $\theta$,
the photon virtuality is given by
	$-q^2 \equiv Q^2 = 4EE'\sin^2\theta/2$,
where the electron mass has been neglected.
The invariant mass squared of the final hadronic state $X$ is
	$W^2 = (p+q)^2 = M^2 + 2 M \nu - Q^2 = M^2 + Q^2(1-x)/x$,
where $M$ is the nucleon mass, $\nu = E-E'$ is the energy transfer,
and $x = Q^2/2M\nu$ is the Bjorken scaling variable.

In the one photon exchange approximation the spin-averaged cross
section for inclusive electron-nucleon scattering in the laboratory
frame can be written as
\begin{equation}
{ d^2\sigma \over d\Omega dE' }
= { \alpha^2 \over Q^4 } {E' \over E} L_{\mu\nu} W^{\mu\nu},
\label{eq:sig_tens}
\end{equation}
where the leptonic tensor
	$L_{\mu\nu}
	= 2 \left( k_\mu k'_\nu + k'_\mu k_\nu
		 - g_{\mu\nu} k \cdot k' \right)$,
and using constraints from Lorentz, gauge and parity invariance
the hadronic tensor $W^{\mu\nu}$ can in general be written as
\begin{equation}
M W^{\mu\nu}
= \left( - g^{\mu\nu} + { q^\mu q^\nu \over q^2} \right)
  F_1(x,Q^2)
+ \left( p^\mu - { p\cdot q \over q^2 } q^\mu \right)
  \left( p^\nu - { p\cdot q \over q^2 } q^\nu \right)
  { F_2(x,Q^2) \over p\cdot q }\, .
\label{eq:Wmunu}
\end{equation}
The structure functions $F_1$ and $F_2$ are generally functions of two
variables, but become independent of the scale $Q^2$ in the Bjorken
limit, in which both $Q^2$ and $\nu \to \infty$ with $x$ fixed.
At finite values of $Q^2$ a modified scaling variable
is more appropriate \cite{Nachtmann,Greenberg},
\begin{equation}
\xi = {2x \over 1 + \rho}\, ,\ \ {\rm with}\
\rho = {|\bm{q}| \over \nu} = \sqrt{1 + {Q^2/\nu^2}}\ ,
\label{eq:xi}
\end{equation}
which tends to $x$ in the Bjorken limit.

In terms of cross sections for absorbing helicity $\pm 1$ (transverse)
and helicity 0 (longitudinal) photons, $\sigma_T$ and $\sigma_L$,
the cross section can be written as
\begin{equation}
\label{eq:gameps}
{ d^2\sigma \over d\Omega dE' }
= \Gamma \left( \sigma_T(x,Q^2) + \epsilon\, \sigma_L(x,Q^2) \right),
\label{eq:sig_phot}
\end{equation}
where $\Gamma = (\alpha/2 \pi^2 Q^2) (E'/E) K/(1-\epsilon)$ is the flux
of transverse virtual photons, with the factor $K = \nu (1-x)$ in the
Hand convention \cite{Hand}, and
\begin{equation}
\epsilon
= \left[ 1 + 2\left( 1+\frac{\nu^2}{Q^2} \right)
	 \tan^2\frac{\theta}{2}
  \right]^{-1}
\end{equation}
is the relative flux of longitudinal virtual photons.
Equating Eqs.~(\ref{eq:sig_tens}) and (\ref{eq:sig_phot}), the structure
functions can be written in terms of the photoabsorption cross sections as
\begin{eqnarray}
\label{eq:f1sigt}
F_1(x,Q^2)
&=& { K \over 4\pi^2\alpha } M \sigma_T(x,Q^2)\, ,       \\
\label{eq:f2sigs}
F_2(x,Q^2)
&=& { K \over 4\pi^2\alpha } { \nu \over (1 + \nu^2/Q^2) }
    \left[ \sigma_T(x,Q^2) + \sigma_L(x,Q^2) \right],
\end{eqnarray} 
which reveals that $F_1$ is related only to the transverse virtual
photon coupling, while $F_2$ is a combination of both transverse and
longitudinal couplings.  One can also define a purely longitudinal
structure function $F_L$,
\begin{equation}
F_L = \rho^2 F_2 - 2xF_1
    = 2xF_1\, R\, ,
\label{eq:fl}
\end{equation}
where $R = \sigma_L/\sigma_T$ is the ratio of longitudinal to
transverse cross sections.

The separation of the unpolarized structure functions into longitudinal
and transverse parts from cross section measurements can be accomplished
via the Rosenbluth, or longitudinal-transverse (LT), separation
technique \cite{Rosen}, by making measurements at two or more values
of $\epsilon$ for fixed $x$ and $Q^2$.  Fitting the reduced cross section
$\sigma/\Gamma$ linearly in $\epsilon$ yields $\sigma_T$ (and therefore
$F_1$) as the intercept, while the ratio $R$ is obtained from the slope.
Note that $F_2$ can only be extracted from cross sections either by
measuring at $\epsilon = 1$ or by performing LT separations.
At typical Jefferson Lab kinematics the contribution of $F_L$ to $F_2$
can be significant.

The above discussion assumed the dominance of the one-photon exchange
amplitude in describing the neutron current electron scattering cross
section.
In principle there are additional contributions arising from the
exchange of a $Z$ boson, and in particular the interference between
$\gamma^*$ and $Z$ exchange.
The interference is in fact very relevant for parity-violating
electron scattering, discussed elsewhere in this volume in connection
with extractions of strange electromagnetic form factors from
parity-violating asymmetries.

\subsection{Operator product expansion}
\label{ssec:ope}

The theoretical basis for describing the $Q^2$ dependence of structure
functions in QCD is Wilson's operator product expansion (OPE)
\cite{OPE}.  The quantities most directly amenable to a QCD analysis
are the {\it moments} of structure functions, the $n$-th moments of
which are defined as
\begin{eqnarray}
M_1^{(n)}(Q^2)     &=& \int_0^1 dx\ x^{n-1} F_1(x,Q^2)\, ,\ \ \ \ \
M_{2,L}^{(n)}(Q^2)\ =\ \int_0^1 dx\ x^{n-2} F_{2,L}(x,Q^2)\, .
\label{eq:mom_def}
\end{eqnarray}
As will become relevant in the discussion of duality in
Sec.~\ref{sec:F2p} below, at large $Q^2 \gg \Lambda_{\rm QCD}^2$ the
moments can be expanded in powers of $1/Q^2$, with the coefficients in
the expansion given by matrix elements of local operators corresponding
to a certain {\em twist}, $\tau$, defined as the mass dimension minus
the spin, $n$, of the operator.  For the $n$-th moment of $F_2$, for
instance, one has the expansion
\begin{eqnarray}
M_2^{(n)}(Q^2) &=&
\sum_{\tau=2,4\ldots}^\infty
{ A_\tau^{(n)}(\alpha_s(Q^2)) \over Q^{\tau-2} }\, ,\ \ \ \
n = 2, 4, 6 \ldots
\label{eq:MnOPE}
\end{eqnarray}
where $A_\tau^{(n)}$ are the matrix elements with twist $\leq \tau$.
As the argument suggests, the $Q^2$ dependence of the matrix elements
can be calculated perturbatively, with $A_\tau^{(n)}$ expressed as a
power series in $\alpha_s(Q^2)$.  For twist two, the coefficients
$A_2^{(n)}$ are given in terms of matrix elements of spin-$n$ operators,
  $A_2^{(n)} p^{\mu_1} \cdots p^{\mu_n} + \cdots
   = \langle p |
     \bar\psi \gamma^{\{\mu_1} iD^{\mu_2} \cdots\, iD^{\mu_n\}} \psi
     | p \rangle$,
where $\psi$ is the quark field, $D^\mu$ is the covariant derivative,
and the braces $\{ \cdots \}$ denote symmetrization of indices and
subtraction of traces.

\begin{figure}[t]
\center\includegraphics[width=12cm]{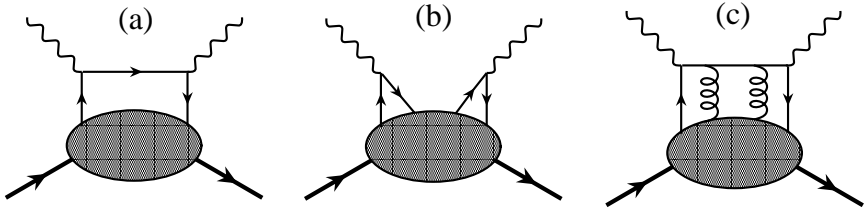}
\caption{{\bf (a)}~Leading-twist (``handbag'') contribution
	to the structure function,
	 {\bf (b)}~higher-twist (``cat's ears'') four-quark
	contributions,
	 {\bf (c)}~higher-twist quark-gluon interactions.}
\label{fig:diag}
\end{figure}

The leading-twist terms correspond to diagrams such as in Fig.~1~(a),
in which the virtual photon scatters incoherently from a single parton.
The higher-twist terms in Eq.~(\ref{eq:MnOPE}) are proportional to
higher powers of $1/Q^2$ whose coefficients are matrix elements of
local operators involving multi-quark or quark-gluon fields, such as
those depicted in Fig.~1~(b) and (c).  The higher twists therefore
parametrize long-distance multi-parton correlations, which can
provide clues to the dynamics of quark confinement.

The additional terms (referred to as the ``trace terms'') in the
twist-two matrix elements involve structures such as
$p^2 g^{\mu_i \mu_j}$ and are thus suppressed by powers of
$p^2/Q^2 \sim Q^2/\nu^2$.  While negligible in the Bjorken limit,
at finite $Q^2$ these give rise to the so-called {\it target mass
corrections} (TMCs), and are important in the analysis of Jefferson Lab
data at large values of $x$.  Because their origin is in the same
twist-two operators that give rise to structure function scaling,
TMCs are formally twist-two effects and are kinematical in origin.
Inverting the expressions for the full moments including the trace
terms, the resulting target mass corrected $F_2$ structure function
in the OPE is given by \cite{GP,TMC}
\begin{equation}
F_2^{\rm TMC}(x,Q^2)
= {x^2 \over \xi^2 \rho^3} F_2^{(0)}(\xi,Q^2)
+ {6 M^2 x^3 \over Q^2 \rho^4}
     \int_\xi^1 du\, 
     \left( 1 + {2 M^2 x \over Q^2 \rho} (u-\xi)
     \right)
{F_2^{(0)}(u,Q^2) \over u^2}\, ,
\label{eq:ope}
\end{equation}
where $F_2^{(0)}$ is the structure function in the $M^2/Q^2 \to 0$ 
limit.  Similar expressions are found for the $F_1$ and $F_L$
structure functions \cite{GP,TMC}.
One should note, however, that the OPE result for TMCs to structure
functions is not unique; in the collinear factorization approach, for
example, in which parton distributions are formulated {\it a priori}
in momentum space, different expressions for TMCs arise \cite{CF}.
While both formalisms give the same results in the Bjorken limit, the 
differences between these at finite $Q^2$ can be seen as representing
an inherent prescription dependence and systematic uncertainty in the 
analysis of structure functions at low $Q^2$.

\section{Proton $F_2$ structure function}
\label{sec:F2p}

\begin{figure}[ht]
\begin{center}
\includegraphics[width=8.1cm]{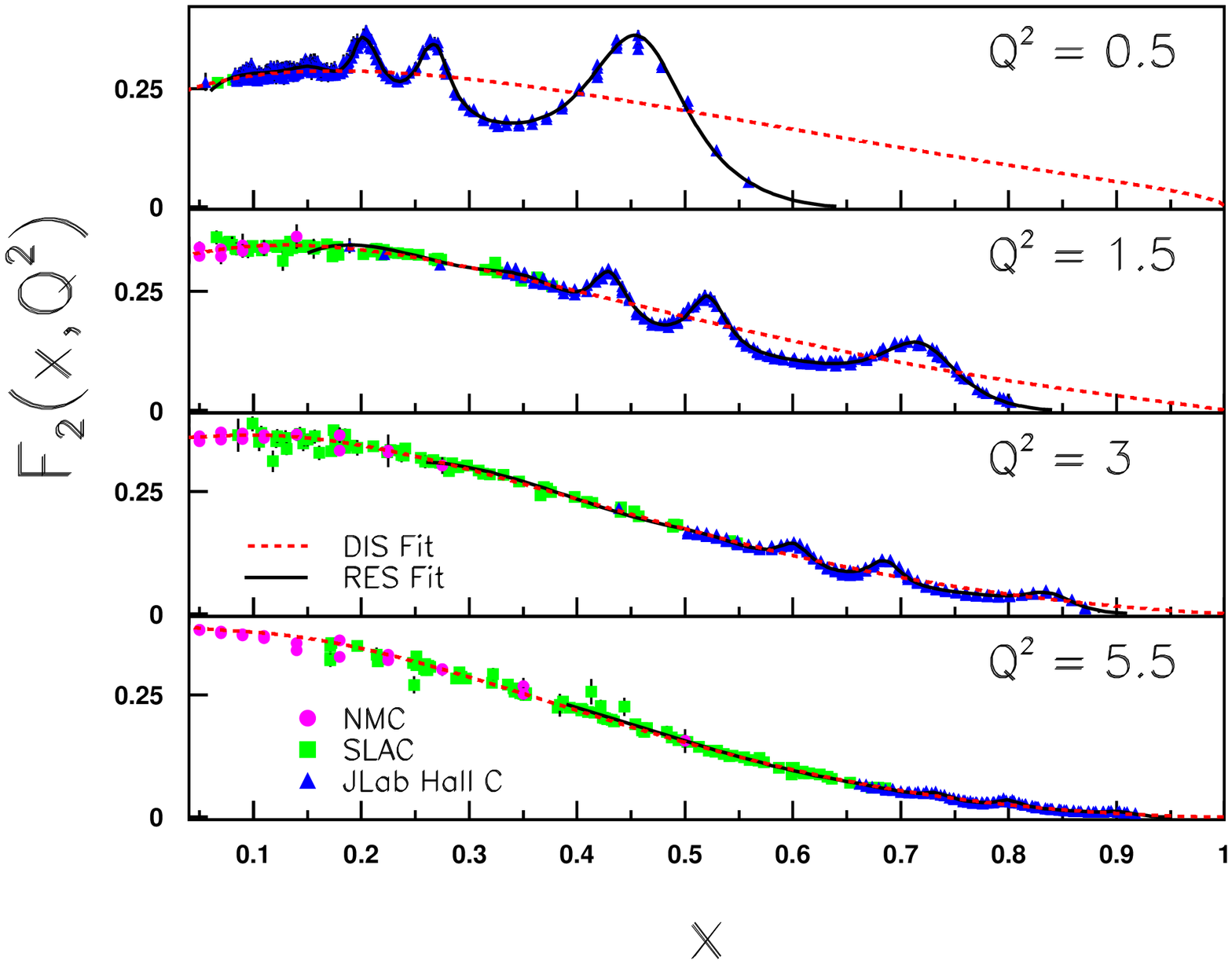} 
\includegraphics[width=6.9cm]{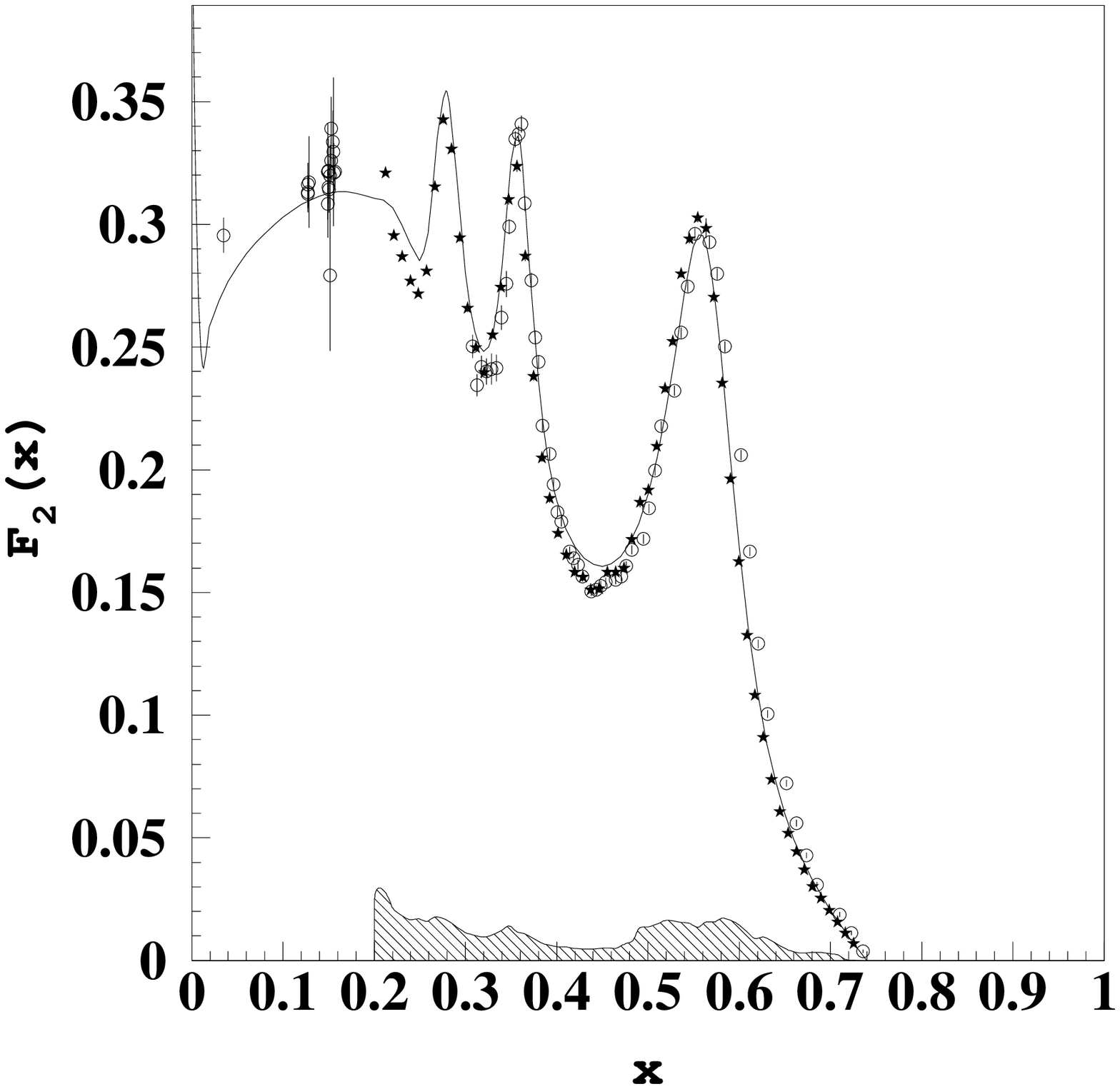}
\caption{\label{fig:f2}
        ({\it Left}) Proton $F_2^p$ structure function data from
        Hall~C \cite{LIANG,SIMONA,e00002}, SLAC \cite{Whitlow},
        and NMC \cite{f2nmc} at $Q^2 = 0.5, 1.5, 3$ and 5.5~GeV$^2$,
	compared with a fit \cite{ResFit} to the transverse and
	longitudinal resonance cross sections from photoproduction
	to $Q^2 = 9$~GeV$^2$ (solid), and a global fit \cite{f2allm}
	to DIS data (dashed).
	({\it Right}) Proton $F_2^p$ from CLAS in Hall~B at
	$Q^2=0.775$~GeV$^2$ (stars) \cite{Osi03}, compared to
	earlier Hall~C data (open circles) \cite{IOANA}.}
\end{center}
\end{figure}

Measurements of the proton $F_2^p$ structure function have been taken
at Jefferson Lab over a range of kinematics, from $Q^2$ as low as 
0.1~GeV$^2$ and below (to study the transition to the photoproduction
point, $Q^2=0$) and up to $Q^2 = 8$~GeV$^2$ (to study the large-$x$
behavior and quark-hadron duality).  At the larger $Q^2$ values the high
luminosity provided by the CEBAF accelerator has allowed significant
improvement in the statistical precision of high-$x$ measurements
over all previous experiments.  In addition, with the HMS spectrometer
in Hall~C well understood, LT separated cross sections have been
measured with better than 1.6\% systematic point-to-point uncertainties
and typically less than 1.8\% normalization uncertainties.

Precision $F_2^p$ spectra extracted from cross sections measured in
Hall~C \cite{LIANG,SIMONA,e00002} are shown in Fig.~\ref{fig:f2}~(left)
as a function of $x$ for several $Q^2$ values ($Q^2 = 0.5, 1.5, 3$
and 5.5~GeV$^2$), together with previous SLAC \cite{Whitlow} and NMC
\cite{f2nmc} data at lower $x$.  The data have been bin-centered to the
common $Q^2$ values shown for all measurements within the range of 20\%
of the central value, utilizing a fit \cite{f2allm} to the DIS data
and a global fit \cite{ResFit} to Jefferson Lab resonance region data.
In addition to the Hall~C measurements, there now also exists a large
body of $F_2^p$ data from Hall~B covering a significant range of
kinematics, which is afforded by the large acceptance of the CLAS
spectrometer.  An example of the $F_2^p$ spectrum extracted from
CLAS is shown in Fig.~\ref{fig:f2}~(right) at $Q^2 = 0.775$~GeV$^2$.
In Table~\ref{tab:jlab_exps} we present a complete list of all
unpolarized inclusive and semi-inclusive measurements on the
proton and deuteron performed at Jefferson Lab through 2010,
including their current status.

\begin{table}[th]
\begin{center}
\caption{Listing of Jefferson Lab unpolarized inclusive and
	semi-inclusive electron-nucleon scattering experiments.}
\label{tab:jlab_exps}
\begin{tabular}{l c c l c l}
\hline
\hline
            &      &         &                              &      	&   	\\
Experiment  & Hall & Target  & Observable                   & Reference	& Status\\
            &      &         &                              &      	&	\\ \hline
            &      &         &                              & 		&	\\
E94-110     & C    & $p$     & $R$ in resonance region	    & \cite{LIANG,94elast,94nonlin,94baldin} & data taken in 1999,     \\
            &      &         & 			            &		& analysis completed \\
            &      &         &                              &   	& \\
E99-118     & C    & $p,d$   & nuclear dependence of        & \cite{e99118} & data taken in 2000, \\
            &      &         & $R$ at low $Q^2$ 	    &		& analysis completed \\
            &      &         &                              &		& \\
CLAS        & B    & $p,d$   & inclusive cross sections     & \cite{Osi03,Osi06} & e1/e2 run periods \\ 
            &      &         &                              &		& \\
E00-002     & C    & $p,d$   & $F_2$ at low $Q^2$           & \cite{e00002} & data taken in 2003, \\
            &      &         &                              &		& analysis completed, \\
            &      &         &                              & 		& pub. in progress \\
            &      &         &                              &		& \\
E00-108     & C    & $p,d$   & semi-inclusive $\pi^\pm$     & \cite{Navasardyan,Mkrtchyan}
									& data taken in 2003, \\
            &      &         & electroproduction            & 		& analysis completed \\
            &      &         &                              &		& \\
E00-116     & C    & $p,d$   & inclusive resonance          & \cite{SIMONA} & data taken in 2003, \\
            &      &         & region cross sections        &		& analysis completed \\
            &      &         & at intermediate $Q^2$	    &		& \\
            &      &         &                              &		& \\
E02-109     & C    & $d$     & $R$ in resonance region	    & \cite{e02109} & data taken in 2005, \\
            &      &         & 				    & 		& analysis in progress \\
            &      &         &                              & 		& \\
E03-012     & B    & $d (n)$ & neutron $F_2^n$ {\it via}    & \cite{BONUS} & data taken in 2005, \\
(BoNuS)     &      &         & spectator tagging            & 		& analysis completed, \\
            &      &         &                              & 		& pub. in progress \\
            &      &         &                              &		& \\
E06-009     & C    & $d$     & $R$ in resonance region	    & \cite{e06009} & data taken in 2007, \\
            &      &         & \& beyond: extension of      & 		& analysis in progress \\
            &      &         & E02-109 to $Q^2=4$~GeV$^2$ &		& \\
            &      &         &                              &		& \\
\hline\hline
\end{tabular}
\end{center}
\end{table}

\subsection{Quark-hadron duality}
\label{ssec:qhd} 

The proton $F_2^p$ data in Fig.~\ref{fig:f2} illustrate the intriguing
phenomenon of quark-hadron duality, which relates structure functions
in the nucleon resonance and DIS regions.  First observed by Bloom and
Gilman \cite{BG} in the early inclusive SLAC data (hence also referred
to as ``Bloom-Gilman duality''), the structure functions in the
resonance region are found on average to equal the structure functions
measured in the ``scaling'' region at higher $W$.  The resonance data 
oscillate around the scaling curve and slide along it with increasing 
$Q^2$, as seen in Fig.~\ref{fig:f2}~(left).

The early $F_2^p$ data from SLAC were extracted from cross sections
assuming a fixed value for $R$ (= 0.18), and with a scaling curve
parametrizing the limited data available in the early 1970's \cite{Mil72}.
Since the original measurements, the $F_2^p$ structure function has
become one of the best studied quantities in lepton scattering, with
data from laboratories around the world contributing to a global data
base spanning over five orders of magnitude in both $x$ and $Q^2$.
With the advent of the Jefferson Lab data, precise $F_2^p$ measurements
now also exist in the resonance region up to $Q^2 \approx 8$~GeV$^2$,
allowing many new aspects of duality to be quantified for the first
time \cite{MEK}.

\begin{figure}
\begin{center}
\includegraphics[height=6.5cm]{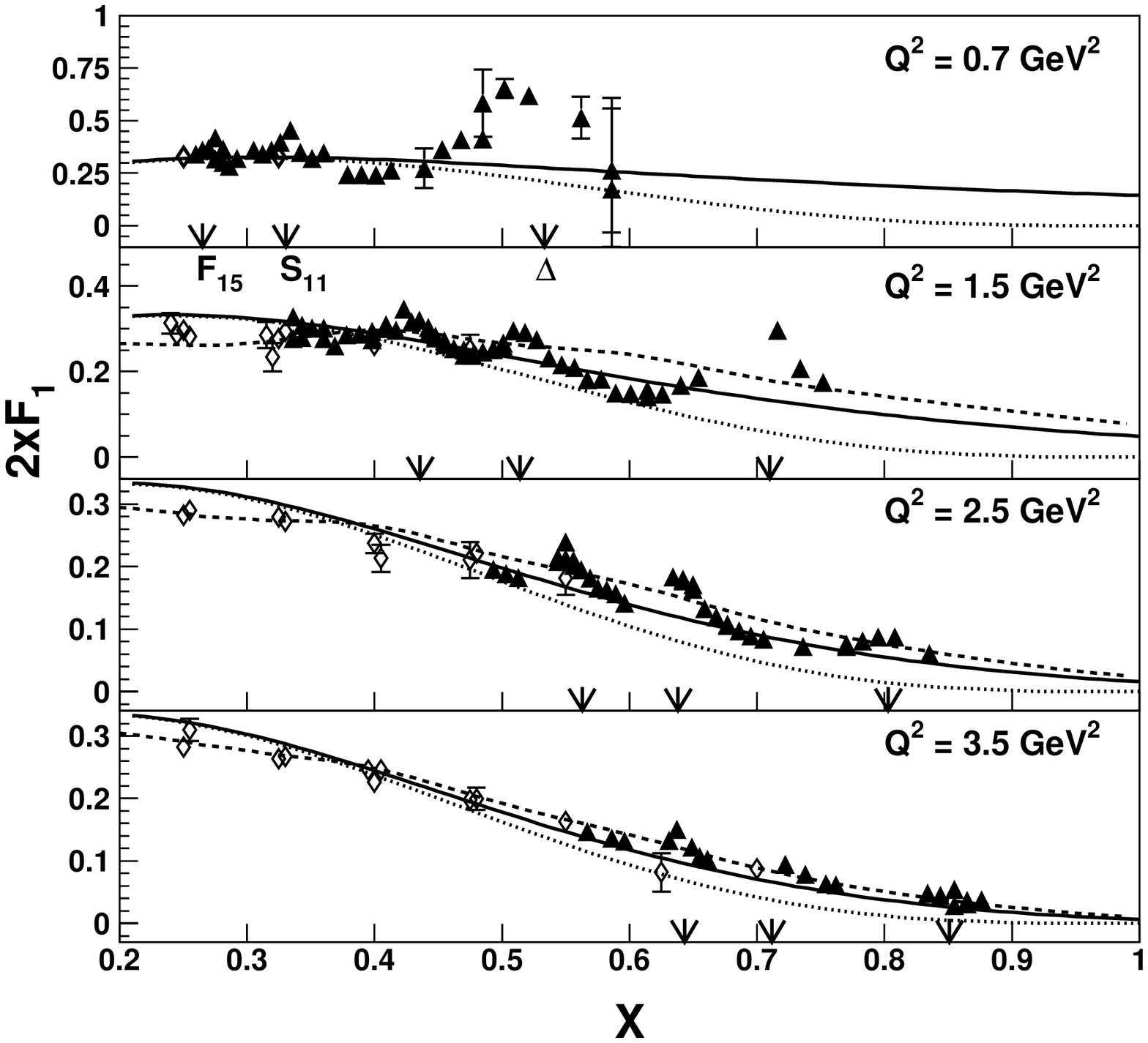}\ \
\includegraphics[height=6.5cm]{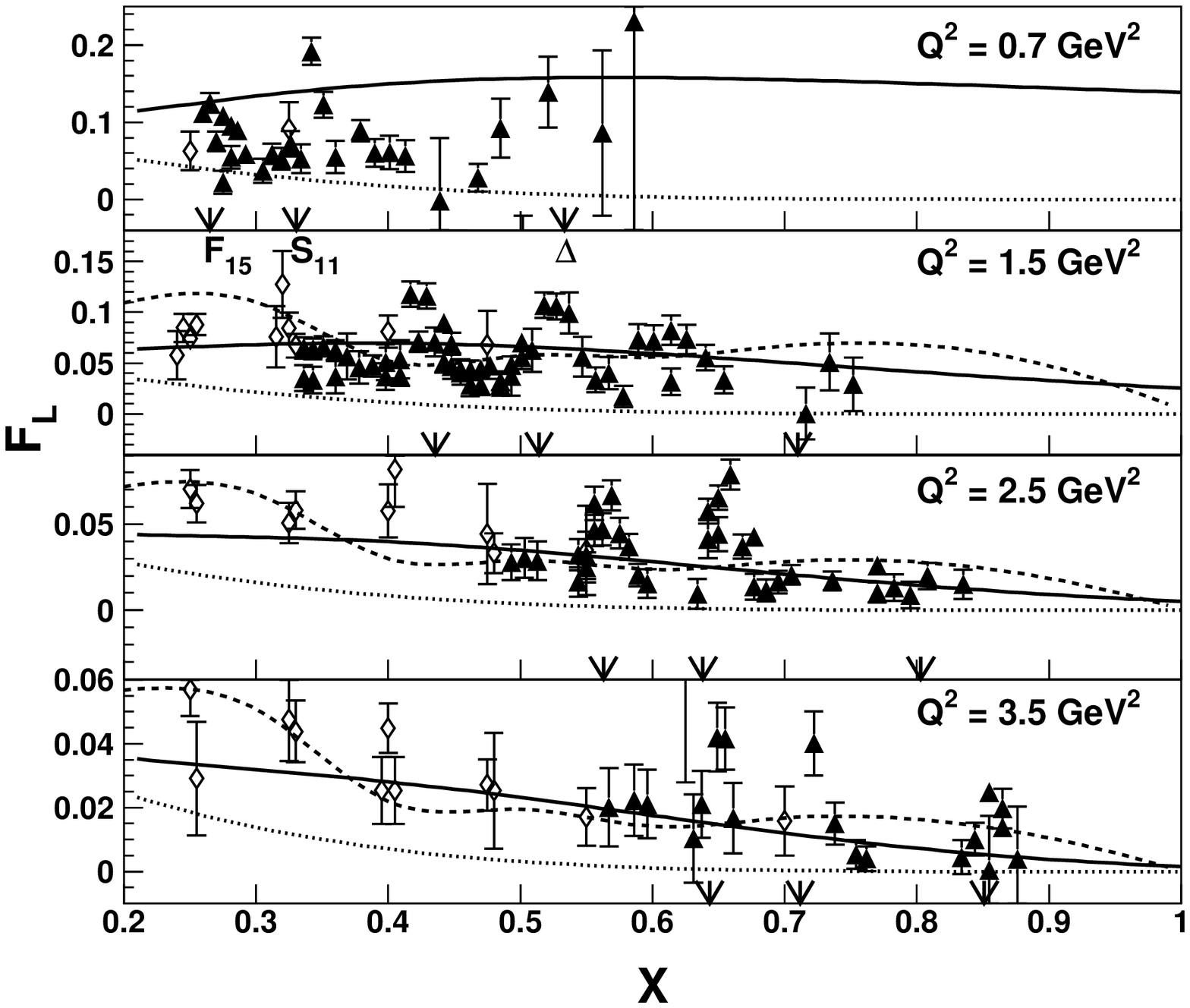}
\caption{Purely transverse {\it (left)} and longitudinal {\it (right)}
	proton structure functions $2xF_1$ and $F_L$ in the resonance
	region \cite{LIANG} (triangles), compared with earlier data
	from SLAC (squares).  The curves are leading twist structure
	functions computed at NNLO from Alekhin (dashed) \cite{Ale03}
	and MRST \cite{MRST04} with (solid) and without (dotted) target
	mass corrections.  The prominent resonance regions
	($\Delta$, $S_{11}$, $F_{15}$) are indicated by the arrows
	along the abscissa.}
\label{fig:F_L}
\end{center}
\end{figure}

While the early duality studies considered only the $F_2$ structure
function \cite{BG}, Jefferson Lab experiments have in addition
revealed the presence of duality in other observables.  For example,
Fig.~\ref{fig:F_L} shows new LT-separated data from Jefferson Lab
experiment E94-110 for the proton transverse ($F_1^p$) and longitudinal
($F_L^p$) structure functions in the nucleon resonance region
\cite{LIANG}.
LT-separated data from SLAC, predominantly in the DIS region, are
also shown for comparison \cite{das94}.  Where they refer to the same
kinematic values, the Jefferson Lab and SLAC data are in excellent
agreement, providing confidence in the achievement of the demanding
precision required of this type of experiment.  In all cases, the
resonance and DIS data merge smoothly with one another in both $x$
and $Q^2$.

The availability of leading twist PDF-based global fits
\cite{CTEQ,MSTW,Alekhin,NNPDF,CTEQ6X} allows comparison of the resonance
region data with leading twist structure functions at the same $x$
and $Q^2$.  The resonance data on the $F_1$ and $F_L$ structure
functions are also found to oscillate around the perturbative QCD
(pQCD) curves, down to $Q^2$ as low as 0.7~GeV$^2$.
Because most of the data lie at large values of $x$ and small $Q^2$,
it is vital for tests of duality to account for the effects of
kinematical target mass corrections \cite{GP}, which give large
contributions as $x \to 1$ \cite{TMC}.
This is clear from Fig.~\ref{fig:F_L}, where the data are compared
with leading twist structure functions computed from PDFs to
next-to-next-to-leading order (NNLO) accuracy from Alekhin \cite{Ale03}
and MRST \cite{MRST04,MRST98}.
The latter are shown with (solid) and without (dotted) target mass
corrections, and clearly demonstrate the importance of subleading
$1/Q^2$ effects at large $x$.  In particular, TMCs give additional
strength at large $x$ observed in the data, which would be significantly
underestimated by the leading twist functions without TMCs.

The phenomenological results raise the question of how can a scaling
structure function be built up entirely from resonances, each of whose
contribution falls rapidly with $Q^2$ \cite{IJMV}?  A number of studies
using various models have demonstrated how sums over resonances can
indeed yield a $Q^2$ independent function (see Ref.~\cite{MEK} for a
review).  The key observation is that while the contribution from each
individual resonance diminishes with $Q^2$, with increasing energy new
states become accessible whose contributions compensate in such a way
as to maintain an approximately constant strength overall.  At a more
microscopic level, the critical aspect of realizing the suppression of
the higher twists is that at least one complete set of even and odd
parity resonances must be summed over for duality to hold \cite{CI}.
For an explicit demonstration of how this cancellation takes place
in the SU(6) quark model and its extensions, see
Refs.~\cite{CI,CM03,CM09}.

\subsection{Structure function moments}
\label{ssec:mom}

The degree to which quark-hadron duality holds can be more precisely
quantified by computing integrals of the structure functions over $x$
in the resonance region at fixed $Q^2$ values,
     $\int_{x_{\rm th}}^{x_{\rm res}}\; dx \; F_2(x,Q^2)$,
where $x_{\rm th}$ corresponds to the pion production threshold at
fixed $Q^2$, and $x_{\rm res} = Q^2/(W_{\rm res}^2 - M^2 + Q^2)$
indicates the $x$ value at the same $Q^2$ where the traditional
delineation between the resonance and DIS regions at
$W = W_{\rm res} \equiv 2$~GeV is made.
These integrals can then be compared to the corresponding integrals
of the structure functions fitted to the higher-$W$, deep-inelastic
data, at the same $Q^2$ and over the same interval of $x$.
The early phenomenological findings \cite{IOANA} suggested that
the integrated strength of the resonance structure functions
above $Q^2 \approx 1$~GeV$^2$ was indeed very similar to that
in the deep-inelastic region, including in each of the individual
prominent resonance regions.
In this section we explore the duality between the resonance and
deep-inelastic structure functions in the context of QCD moments.

According to De~Rujula, Georgi and Politzer \cite{DGP}, one can
formally relate the appearance of quark-hadron duality to the
vanishing suppression of higher twist matrix elements in the
QCD moments of the structure functions \cite{OPE}.
Namely, if certain moments of structure functions are observed
to be independent of $Q^2$, as implied by duality, then from
Eq.~(\ref{eq:MnOPE}) the moments must be dominated by the leading,
$Q^2$ independent term, with the $1/Q^{\tau-2}$ higher twist terms
suppressed.
Duality is then synonymous with the suppression of higher twists,
which in partonic language corresponds to the suppression, or
cancellation, of interactions between the scattered quark and the
spectator system such as those illustrated in Fig.~1~(b) and (c).
Conversely, if the moments display power-law $Q^2$ dependence, then
this implies violation of duality; moreover, if the violation is not
overwhelming, the $Q^2$ dependence of the data can be used to extract
information on the higher twist matrix elements \cite{JiUnrau}.

\begin{figure}[hb]
\includegraphics[width=10cm]{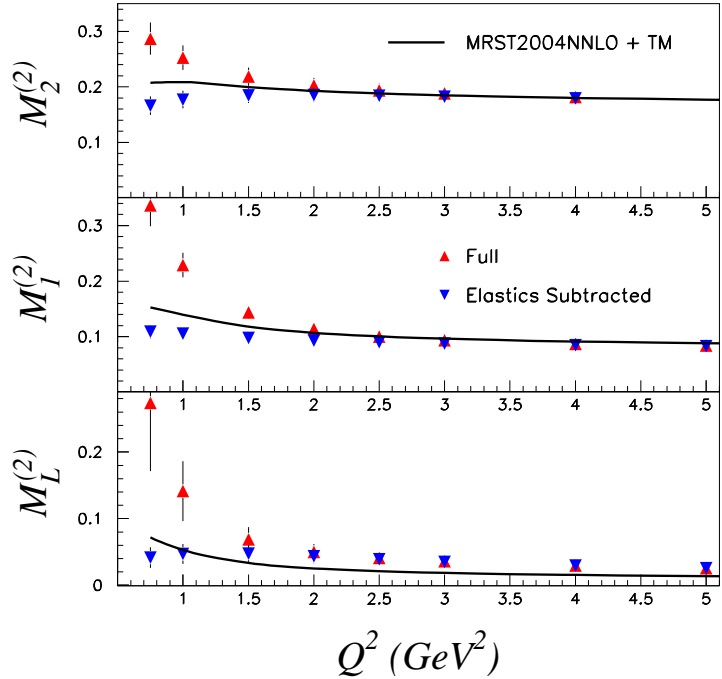}
\begin{centering}
\caption{Total $n=2$ moments of the proton $F_2^p$ {\it (top)},
	$F_1^p$ {\it (center)} and $F_L^p$ {\it (bottom)} structure
	functions determined from global fits to existing DIS data
	and Jefferson Lab resonance region data \cite{Chr04},
	compared with moments computed from the leading twist
	PDFs from MRST at NNLO \cite{MRST04}.}
\label{fig:f2ht}
\end{centering} 
\end{figure}

The first determination \cite{Arm01} of the $F_2^p$ moments from
Jefferson Lab data was made utilizing structure functions measured
in Hall~C \cite{IOANA}, while a later evaluation \cite{Osi03}
included the large body of additional data from CLAS.
More recently, the extraction of $F_2^p$ has been further enhanced
by LT-separated data from Hall~C, shown in Fig.~\ref{fig:f2} along
with the fit \cite{ResFit} to the LT separated cross sections.
The $n=2$ moments for the proton $F_2^p$, $F_1^p$, and $F_L^p$
structure functions are shown in Fig.~\ref{fig:f2ht} versus $Q^2$,
as determined from integrating this fit.  For $F_2^p$ they are
found to be in very good agreement with the earlier measurements.
Also shown is the leading-twist contribution calculated from the
MRST parameterization \cite{MRST04}, corrected for target mass
effects \cite{GP}.

One of the most striking features of the results in Fig.~\ref{fig:f2ht} 
is that the elastic-subtracted moments exhibit the same $Q^2$ dependence 
as the PDF fits down to $Q^2 \approx 1$~GeV$^2$.  Even with the elastic
contribution included, which vanishes in the Bjorken limit and is hence
pure higher twist, there is excellent agreement between the resonance
and DIS data for $Q^2 \gtorder 2$~GeV$^2$.
Until very recently \cite{Alekhin,CTEQ6X}, this fact has not been
widely appreciated or utilized in global PDF fitting efforts
\cite{CTEQ,MSTW}, which typically impose cuts on data of
$Q^2 \gtorder 4$~GeV$^2$ and $W^2 \gtorder 12$~GeV$^2$.

While the OPE provides a systematic approach to identifying and
classifying higher twists, it does not reveal {\it why} these
are small or {\it how} duality is realized globally and locally.
To further explore the {\it local} aspects of duality within a
perturbative QCD context, a ground-breaking new approach using
``truncated'' moments of structure functions, developed recently by
Forte {\it et al.} \cite{Forte} and extended by Kotlorz \& Kotlorz
\cite{Kotlorz}, was applied by Psaker {\em et al.} \cite{Ales} to
Jefferson Lab data.
The virtue of truncated moments is that they obey a similar set
of $Q^2$ evolution equations as those for PDFs themselves, which
therefore enables a rigorous connection to be made between local
duality and QCD.
It allows one to quantify the higher twist content of various
resonance regions, and determine the degree to which individual
resonances are dominated by the leading twist components.

Defining the $n$-th truncated moment ${\cal M}_n$ of a PDF $q(x,Q^2)$
between $x_{\rm min}$ and $x_{\rm max}$ as
\begin{eqnarray}
{\cal M}_n(x_{\rm min},x_{\rm max},Q^2)
&=& \int_{x_{\rm min}}^{x_{\rm max}} dx\;
    x^{n-1}\ q(x,Q^2)\ ,
\label{eq:trunc_def}
\end{eqnarray}
the evolution equations for the truncated moments can be written as
\begin{eqnarray}
\frac{d{\cal M}_n}{dt}
& = & \frac{\alpha_s(Q^2)}{2\pi}
\left( P'_n \otimes {\cal M}_n \right)\ ,\ \ \ \ 
t = \ln\left(Q^2/\Lambda_{\rm QCD}^2\right)\, .
\label{eq:trunc_evol}
\end{eqnarray}
The symbol $\otimes$ denotes the Mellin convolution of the truncated
moment and the ``splitting function'' $P'_n$, which is related to the
usual DGLAP evolution splitting function $P$ \cite{DGLAP} by
$P'_n(z,\alpha_s(Q^2)) = z^n\ P(z,\alpha_s(Q^2))$.
The extent to which nucleon structure function data in specific regions
in $x$ (or $W$) are dominated by leading twist can be determined by
constructing empirical truncated moments and evolving them to different
$Q^2$.
Deviations of the evolved moments from the experimental data at the new
$Q^2$ then reveal any higher twist contributions in the original data.

A next-to-leading order (NLO) analysis \cite{Ales} of data on the proton
$F_2^p$ structure function from Jefferson Lab covering a range in $Q^2$
from 1~GeV$^2$ to $\approx$~6~GeV$^2$ revealed intriguing behavior
of the higher twists for different nucleon resonance regions.
Assuming that $F_2^p$ data beyond a large enough $Q^2$ (taken to be
$Q^2 = Q_0^2 = 25$~GeV$^2$ in Ref.~\cite{Ales}) are dominated by
leading twist, the truncated moments were computed at $Q_0^2$ and
evolved to lower $Q^2$.
Note that the truncated moments are computed over the range
$W_{\rm th} \leq W \leq W_{\rm max}$, where the
$W_{\rm th} = M + m_\pi$ is the inelastic threshold.
At $Q^2 = 1$~GeV$^2$ this corresponds to the integration range
$x_{\rm min} \leq x \leq x_{\rm th}$, where
$x_{\rm th} = \left[ 1 + m_\pi (m_\pi+2M)/Q^2\right]^{-1} \simeq 0.78$.

\begin{figure}[ht]
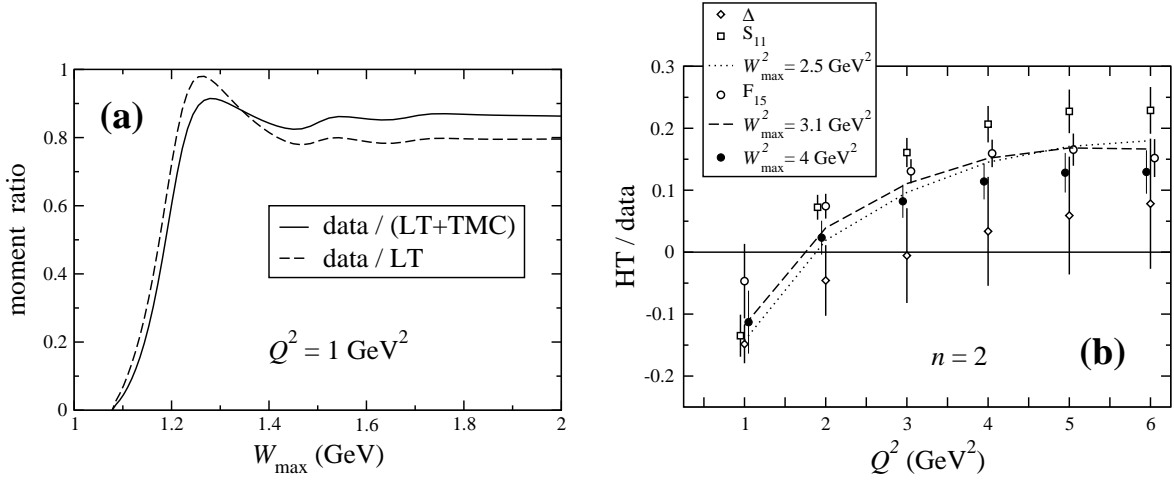

\begin{center}
\includegraphics[scale=0.3]{fig5a.eps}\hspace*{0.5cm}
\includegraphics[scale=0.3]{fig5b.eps}
\end{center}
\caption{{\bf (a)}
	Ratio of the ${\cal M}_2$ truncated moments of the data
	to the leading twist + TMC (solid), and data to leading
        twist without TMC (dashed) at $Q^2 = 1$~GeV$^2$, as a
	function of $W_{\rm max}$.
	{\bf (b)}
	$Q^2$ dependence of the fractional higher twist (HT)
        contribution to the $n=2$ truncated moment data,
	for various intervals in $W$.}
\label{fig:trunc}
\end{figure}

The ratio of the truncated moments of the data to the leading twist
is shown in Fig.~\ref{fig:trunc}(a) as a function of $W_{\rm max}$
at $Q^2 = 1$~GeV$^2$, with and without target mass corrections.
Including the effects of TMCs, the leading twist moment differs
from the data by $\sim 15\%$ for $W_{\rm max} > 1.5$~GeV.
To quantify the higher twist content of the specific resonance
regions, and at different values of $Q^2$, several intervals in
$W$ are considered:
$W_{\rm th}^2 \leq W^2 \leq$~1.9~GeV$^2$ ($\Delta(1232)$ or first
resonance region);
$1.9 \leq W^2 \leq 2.5$~GeV$^2$ ($S_{11}(1535)$ or second resonance
region); and
$2.5 \leq W^2 \leq 3.1$~GeV$^2$ ($F_{15}(1680)$ or third resonance
region).
The higher twist contributions to ${\cal M}_2$ in these regions are
shown in Fig.~\ref{fig:trunc}(b) as ratios to moments of the data.

The results indicate deviations from leading twist behavior of the 
entire resonance region data (filled circles in Fig.~\ref{fig:trunc}(b))
at the level of $\ltorder 15\%$ for all values of $Q^2$ considered,
with significant $Q^2$ dependence for $Q^2 \ltorder 4$~GeV$^2$.
The strong $Q^2$ dependence of the higher twists is evident here in
the change of sign around $Q^2 = 2$~GeV$^2$, with the higher twists
going from $\approx -10\%$ at $Q^2 = 1$~GeV$^2$ to $\approx 10$--15\%
for $Q^2 \sim 3-6$~GeV$^2$.
At larger $Q^2$ the higher twists are naturally expected to decrease,
once the leading twist component of the moments begins to dominate.

Interestingly, the magnitude of the higher twist contributions in the
$\Delta$ region (diamonds) is smallest, decreasing from $\approx -15\%$
of the data at $Q^2=1$~GeV$^2$ to values consistent with zero at larger
$Q^2$.
The higher twists are largest, on the other hand, for the $S_{11}$
region (squares), where they vary between $\approx -15\%$ of the   
data at $Q^2=1$~GeV$^2$ and 20--25\% at $Q^2 \sim 5$~GeV$^2$.
Combined, the higher twist contribution from the first two resonance
regions (dotted curve) is $\ltorder 15\%$ in magnitude for all $Q^2$.
The rather dramatic difference between the $\Delta$ and the $S_{11}$, 
may, at least in part, be due to the choice of the differentiation  
point of $W^2 = 1.9$~GeV$^2$.
A lower $W^2$ choice, for instance, would lower the higher twist
content of the $S_{11}$ at large $Q^2$, while raising that of the
$\Delta$.
However, this $W^2$ choice corresponds to the local minimum between 
these two resonances in the inclusive spectra, and is the one most 
widely utilized.

The higher twist content of the $F_{15}$ region (open circles) is
similar to the $S_{11}$ at low $Q^2$, but decreases more rapidly
for $Q^2 > 3$~GeV$^2$.
The higher twist content of the first three resonance regions
combined (dashed curve) is $\ltorder 15$--20\% in magnitude for
$Q^2 \leq 6$~GeV$^2$.
Integrating up to $W_{\rm max}^2 = 4$~GeV$^2$ (filled circles),
the data on the $n=2$ truncated moment are found to be leading twist
dominated at the level of 85--90\% over the entire $Q^2$ range.

The results in Fig.~\ref{fig:trunc} can be compared with quark model
expectations \cite{CI,CM03}, which predict systematic deviations of
resonance data from local duality.
Assuming dominance of magnetic coupling, the proton data are expected
to overestimate the DIS function in the second and third resonance
regions due to the relative strengths of couplings to odd-parity
resonances; the positive higher twists observed in 
Fig.~\ref{fig:trunc}(b) for $Q^2 \gtorder 2$~GeV$^2$ indeed support
these predictions.

\section{Deuteron $F_2$ measurements}
\label{sec:F2d}

Together with hydrogen, inclusive lepton scattering from deuterium
targets has provided an extensive data base of $F_2$ structure
function measurements over a large range of kinematics.
While a significant quantity of $F_2^d$ data was collected from
experiments at CERN and SLAC, the quasi-elastic and nucleon
resonance regions, especially at low and moderate $Q^2$
($\approx$ few GeV$^2$), were only mapped precisely with the
advent of Jefferson Lab data \cite{IOANA,Osi06}.
Similarly, Jefferson Lab contributed with precision $F_2^d$ data
in the region of $x > 1$ \cite{Arr01}, of relevance for constructing
higher moments of deuteron structure functions.

Experiments in Hall~C \cite{IOANA,Arr01} have provided $F_2^d$ data
in select regions of $x$ and $Q^2$, while inclusive cross section   
measurements in CLAS have covered a continuous two-dimensional region
over the entire resonance region up to $Q^2 = 6$~GeV$^2$.
This combination is rather useful for determining moments of $F_2$.
In such extractions one usually assumes that the ratio $R$ for the
deuteron is similar to that for the proton at scales $Q^2$ of a
few GeV$^2$ \cite{Tva07}.
New measurements which will test this assumption will be reviewed
in Sec.~\ref{sec:FL}.

\begin{table}[ht]
\caption{Lowest two moments (for $n=2$ and 4) of the isovector
	$F_2$ structure function.  Experimental results for $Q^2
	\approx 4$~GeV$^2$ are compared with lattice calculations
	extrapolated to the chiral limit.\\}
\begin{center}
\begin{tabular}{|c|c|c|c|}			\hline \hline
$n$ & Niculescu {\it et al.} \cite{NAEK06}
	        & Osipenko {\it et al.} \cite{OMSKR06}
			     & Detmold {\it et al.} \cite{DMT02} \\
    & (Hall C)  & (Hall B)   & (lattice)		\\ \hline
 2  & 0.049(17) & 0.050(9)   & 0.059(8) \\
 4  & 0.015(3)  & 0.0094(16) & 0.008(3) \\ \hline \hline 
\end{tabular}
\label{tab:momcomp}
\end{center}
\end{table}

The results of the moment analyses are shown in Table~\ref{tab:momcomp},
expressed as the isovector (proton minus neutron) combination, and
compared with isovector moments from lattice QCD \cite{DMT02,Det01}.
For simplicity the neutron moments here are defined as the difference
between the deuteron and proton moments --- see, however,
Sec.~\ref{ssec:F2n_extract} below.  The $n=2$ moments from the Hall~B
\cite{OMSKR06} and Hall~C \cite{NAEK06} analyses agree well with each
other, and with the lattice extraction, which includes the effects of
pion loops and the intermediate $\Delta$(1232) resonance in the chiral
extrapolation.
For the $n=4$ moment the comparison between the Hall~B and Halls~C
results shows a slight discrepancy, which may be reduced once
precision high-$x$ data at $Q^2 \sim 4$~GeV$^2$ are included in
the extractions.

Analysis of the $Q^2$-dependence of the deuteron $F_2$ moments in
Ref.~\cite{Osi06} suggests a partial cancellation of different higher
twist contributions entering in the OPE with different signs, which
is one of the manifestations of quark-hadron duality.
The slow variation with $Q^2$ of the structure function moments,
down to $Q^2 \approx 1$~GeV$^2$, was also found in the analysis of
proton data \cite{Arm01}, where such cancellations were found to be
mainly driven by the elastic contribution.
Furthermore, by comparing the proton and (nuclear corrected) neutron
structure function moments, the higher twist contributions were found
to be essentially isospin independent \cite{Osi07}.  This suggests
the possible dominance of $ud$ correlations over $uu$ and $dd$ in
the nucleon, and implies higher twist corrections that are consistent
with zero in the isovector $F_2$ structure function.

More recently, high precision deuteron cross sections in the resonance
region have been measured in Hall~C \cite{e02109,e06009} with the aim
of providing LT separated deuteron structure functions of comparable 
precision and kinematic coverage to those performed for the proton.  
These higher precision LT separated data will allow for a significant
further reduction in the uncertainties in the current nonsinglet $F_2$
extractions.

\section{Neutron $F_2$ structure function}
\label{sec:F2n} 

A complete understanding of the valence quark structure of the nucleon
requires knowledge of both its $u$ and $d$ quark distributions.
While the $u$ distribution is relatively well constrained by
measurements of the proton $F_2^p$ structure function, in contrast
the $d$ quark distribution is poorly determined due to the lack of
comparable data on the neutron structure function $F_2^n$.
The absence of free neutron targets makes it necessary to use light
nuclei such as deuterium as effective neutron targets, and one must
therefore deal with the problem of extracting neutron information
from nuclear data.

\subsection{Neutron structure from inclusive $F_2$ data}
\label{ssec:F2n_extract}

In standard global PDF analyses, sensitivity to the $d$-quark from
charged lepton scattering is primarily provided by the neutron in the
deuteron.  Usually the neutron $F_2^n$ structure function is extracted
by subtracting the deuteron and proton structure function data assuming
that nuclear corrections are negligible.  At large $x$, however, the
ratio of the deuteron to free nucleon structure functions is predicted
to deviate significantly from unity \cite{MST,KPW,KP06,KMK}, which can
have significant impact on the behavior of the extracted neutron
structure function at large $x$ \cite{CTEQ6X,MT}.

Even when nuclear effects are considered, there exist practical
difficulties with extracting information on the free neutron from
nuclear data, especially in the nucleon resonance region, where
resonance structure is largely smeared out by nucleon Fermi motion.
A recent analysis \cite{MKMK} used a new method \cite{KMK} to extract
$F_2^n$ from $F_2^d$ and $F_2^p$ data, in which nuclear effects are
parameterized via an additive correction to the free nucleon structure
functions, in contrast to the more common multiplicative method
\cite{BODEK} which fails for functions with zeros or with non-smooth
data.
In the standard impulse approximation approach to nuclear structure
functions, the deuteron structure function can be written as a
convolution \cite{KP06,KMK}
\begin{eqnarray}
F_2^d(x,Q^2)
&=& \sum_{N=p,n}
    \int dy\ f_{N/d}(y,\rho)\ F_2^N(x/y,Q^2)\, ,
\label{eq:conv_def}
\end{eqnarray}
where $f_{N/d}$ is the light-cone momentum distribution of nucleons
in the deuteron (or ``smearing function''), and is a function of the
momentum fraction $y$ of the deuteron carried by the struck nucleon,
and of the virtual photon ``velocity'' $\rho$ (see Eq.~(\ref{eq:xi})).
The smearing function encodes the effects of the deuteron wave function,
accounting for nuclear Fermi motion and binding effects, as well as
kinematical finite-$Q^2$ corrections.
Although not well constrained, nucleon off-shell effects have also been
studied \cite{MST,KPW,KP06}; their influence appears to be small compared
with the errors on the existing data, except at very large $x$.

\begin{figure}[ht]
\begin{center}  
\vspace*{1cm}
\includegraphics[width=7.5cm,height=5cm]{fig6a.eps}\hspace*{0.5cm}
\includegraphics[width=7.5cm,height=5cm]{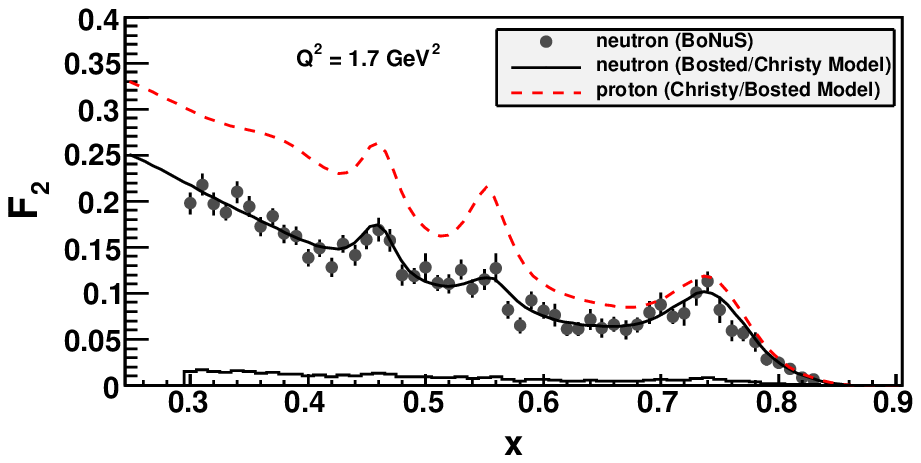}
\end{center}
\caption{{\it (Left)}
	Neutron $F_2^n$ structure function extracted from inclusive
	deuteron and proton data at $Q^2=1.7$~GeV$^2$ \cite{MKMK},
	together with the reconstructed $F_2^d$.
	{\it (Right)}
	Neutron structure function extracted from the BoNuS
	experiment \cite{BONUS}, compared with the Bosted/Christy
	model of the neutron $F_2^n$ \cite{ResFitd} and the
	corresponding Christy/Bosted parametrization for $F_2^p$
	\cite{ResFit}.}
\label{fig:F2n}
\end{figure}

In Fig.~\ref{fig:F2n} we illustrate the results of a typical
extraction of $F_2^n$ from Jefferson Lab $F_2^d$ and $F_2^p$
data at $Q^2 = 1.7$~GeV$^2$.
The proton data show clear resonant structure at large $x$, which
is mostly washed out in the deuteron data.  The resulting neutron
$F_2^n$ is shown after two iterations of the procedure with an
initial guess of $F_2^n=F_2^p$.  Clear neutron resonance structure
is visible in the first ($\Delta$) and second resonance regions,
at $x \sim 0.75$ and 0.55, respectively, with some structure visible
also in the third resonance region at $x \sim 0.45$, albeit with
larger errors.
The deuteron $F_2^d$ reconstructed from the proton and extracted
neutron data via Eq.~(\ref{eq:conv_def}) indicates the relative
accuracy and self-consistency of the extracted $F_2^n$.

Of course it is not possible to avoid nuclear model dependence in the
inversion procedure, and some differences in the extracted $F_2^n$
will arise using different models for the smearing functions $f_{N/d}$.
To remove, or at least minimize, the model dependence in the extracted 
free neutron structure, several methods have been proposed, such as
utilizing inclusive DIS from $A=3$ mirror nuclei
\cite{Afnan,Salme,MARATHON}, and semi-inclusive DIS from a deuteron
with spectator tagging \cite{BONUS,MSS}.  In addition, experiments
involving weak interaction probes \cite{PVDIS,SOLID,HM} can provide the
information on the flavor separated valence quark distributions directly.
In the next section we discuss in detail one of these methods for
determining the free neutron structure, namely the BoNuS experiment
at Jefferson Lab \cite{BONUS}.

\subsection{Tagged neutron structure functions}
\label{ssec:F2n_bonus}

To overcome the absence of free neutron targets, the Hall~B BoNuS
(Barely Off-shell NeUtron Structure) experiment has measured inclusive
electron scattering on an almost free neutron using the CLAS
spectrometer and a recoil detector to tag low momentum protons.
The protons are tracked in a novel radial time projection chamber
\cite{bonus-nim} utilizing gas electron multiplier foils to amplify the
proton ionization in a cylindrical drift region filled with a mixture of
Helium and Di-Methyl Ether, and with the proton momentum determined from
the track curvature in a solenoidal magnetic field.
Slow backward-moving spectator protons are tagged with momenta as low
as 70~MeV in coincidence with the scattered electron in the reaction
$D(e,e' p_s)X$.
This technique ensures that the electron scattering took place on an
almost free neutron, with its initial four-momentum inferred from the
observed spectator proton.  Cutting on spectator protons with momenta
between 70 and 120~MeV and laboratory angles greater than 120~degrees
minimizes the contributions from final state interactions and off-shell
effects to less than several percent on the extracted neutron cross
section.

While inclusive scattering from deuterium results in resonances which
are significantly broadened (often to the point of being unobservable),
determination of the initial neutron momentum allows for a dramatic
reduction in the fermi smearing effects and results in reconstructed
resonance widths comparable to the inclusive proton measurements.
In addition, the large CLAS acceptance for the scattered electron
allowed for the tagged neutron cross section to be measured over a
significant kinematic range in both $W^2$ and $Q^2$ at beam energies
of 4.2~GeV and 5.3~GeV.  BoNuS $F_2^n$ data extracted at a beam energy
of 5.3~GeV and $Q^2 = 1.7~\rm GeV^2$ are shown in Fig.~\ref{fig:F2n}
(right panel) for $x$ values from pion production threshold through
the resonance region and into the DIS regime.  This will allow for the
first time a unambiguous study of the inclusive neutron resonance
structure functions.

An extension of BoNuS has been approved \cite{BONUS12} to run at a
beam energy of 11~GeV after the energy upgrade of the CEBAF accelerator.
The kinematic coverage will allow the extraction of the ratio of neutron
to proton structure functions $F_2^n/F_2^p$ to $x$ as large as 0.8,
and the corresponding $d$ to $u$ large x parton distribution ratio.
Other quantities which BoNuS will make possible to measure include the
elastic neutron form factor, quark-hadron duality on the neutron,   
semi-inclusive DIS and resonance production channels, hard exclusive
reactions such as deeply-virtual Compton scattering or deeply-virtual 
meson-production from the neutron, as well as potentially the inclusive
structure function of a virtual pion.

\section{Longitudinal structure function $F_L$}
\label{sec:FL}

The unpolarized inclusive proton cross section contains two
independent structure functions.  While $F_2^p$ has been measured to
high precision over many orders of magnitude in both $x$ and $Q^2$,
measurements of the longitudinal structure function $F_L^p$ (and the
ratio $R$) have been significantly more limited in both precision and
kinematic coverage.  This is due in part to the challenges inherent
in performing LT separations, which typically require point-to-point
systematic uncertainties in $\epsilon$ to be smaller than 2\% to
obtain uncertainties on $F_L$ of less than 20\%.

While the inclusive cross section is proportional to
$2xF_1 + \epsilon F_L$, the $F_2$ structure function $\sim 2xF_1 + F_L$,
so that $F_2$ is only proportional to the cross section for
$\epsilon=1$.  At high $Q^2$ the scattering of longitudinal
photons from spin-1/2 quarks is suppressed, and in the parton
model one expects $F_L$ (and $R$) to vanish as $Q^2 \to \infty$.
At low $Q^2$, however, $F_L$ is no longer suppressed, and could
be sizable, especially in the resonance region and at large $x$.
On the other hand, $F_L$ is dominated by the gluon contribution
at small $x$, where new measurements from HERA \cite{herafl}
have shown that it continues to rise.  In the kinematic range
of Jefferson Lab $F_L$ has been found to be typically 20\% of
the magnitude of $F_2$, which is consistent with earlier SLAC
measurements where the kinematic regions overlap.

\begin{figure}[ht]
\begin{center}
\includegraphics[width=7cm,height=7.5cm]{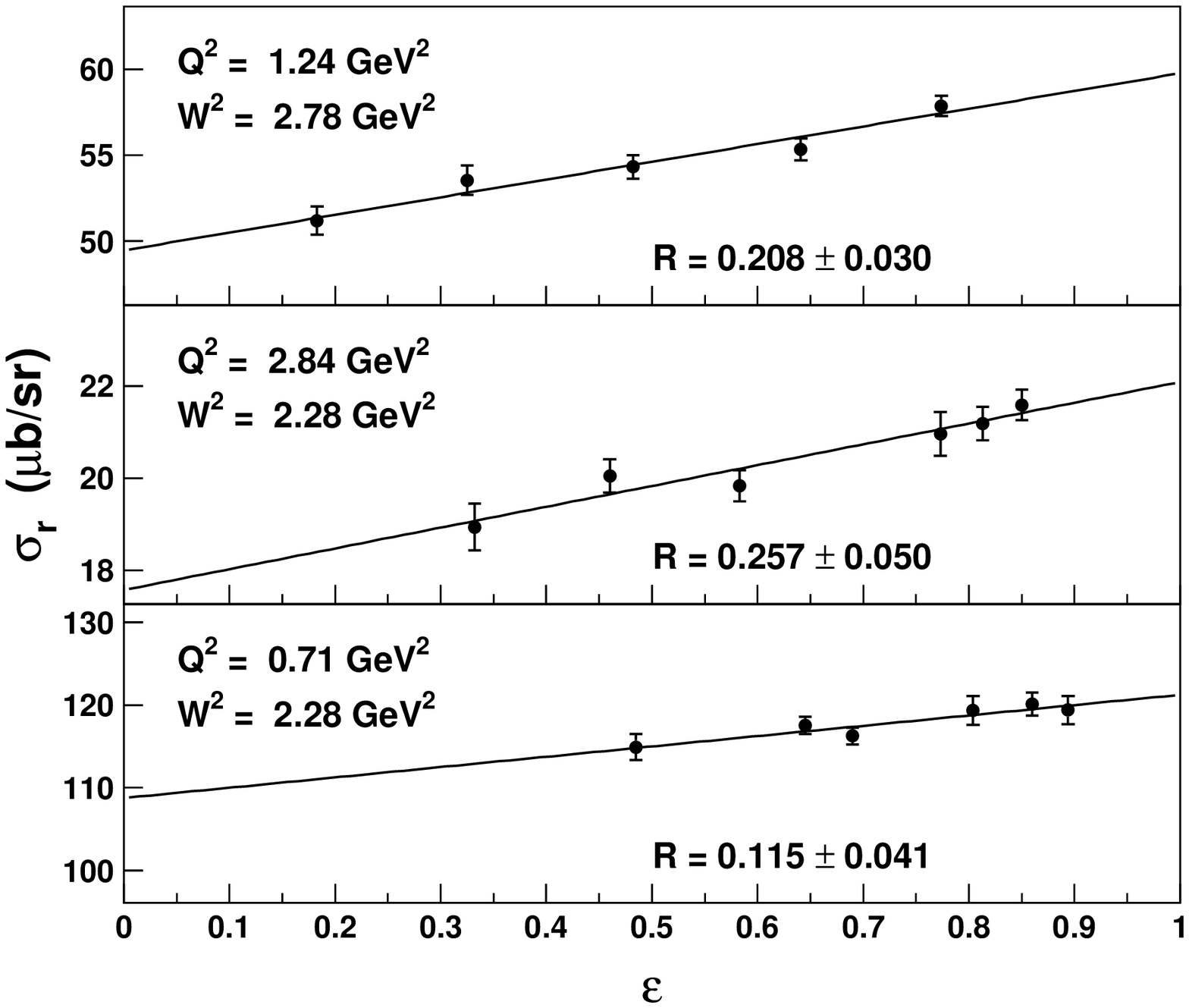}
\includegraphics[width=8.5cm,height=8.5cm]{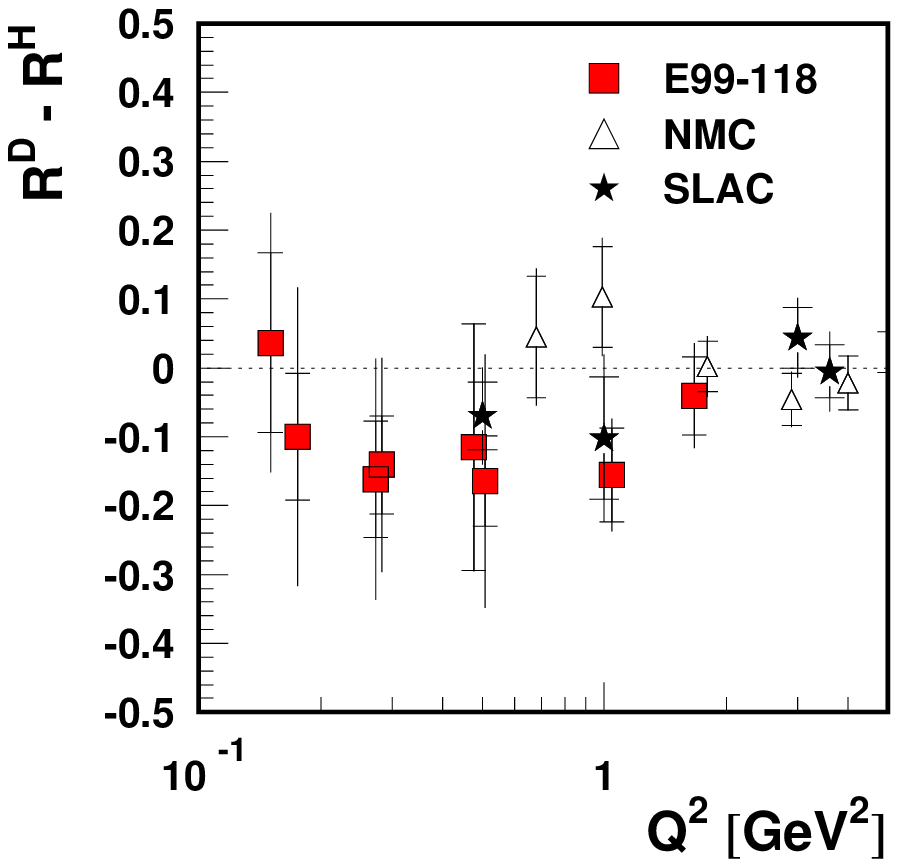}
\end{center}
\caption{{\it (Left)} Sample LT separations from Jefferson Lab
	experiment E94-110 \cite{LIANG}.
	 {\it (Right)} Difference between the ratios $R$ in
	deuterium and hydrogen versus $Q^2$, from E99-118
	\cite{TvaskisLT}, compared with previous NMC and SLAC
	measurements.}
\label{fig:rd-rp}
\end{figure}

An extensive program of LT separations has been carried out in Hall~C,
including measurements of the longitudinal strength in the resonance
region for $0.3 < Q^2 < 4.5$~$\rm GeV^2$ for both proton \cite{LIANG}
and deuteron \cite{e02109,e06009} targets.
The Jefferson Lab experiments listed in Table~\ref{tab:jlab_exps}
for which LT separations have been performed for the proton or 
deuteron are E94-110, E99-118, E00-002, E02-109, and E06-009.
The Jefferson Lab data complement well the previous results at
smaller $x$ from SLAC and NMC in this $Q^2$ region, and improve
dramatically on the few measurements that existed below
$Q^2 = 8$~GeV$^2$ in this $x$ region, which had typical
errors on $R$ and $F_L$ of 100\% or more
--- see Fig.~\ref{fig:rd-rp} (left).

The recent precision LT separated measurements of proton cross
sections \cite{LIANG} have allowed for the first time detailed
duality studies in all of the unpolarized structure functions
and their moments.
The results of the proton separated structure functions in
the resonance region were presented in Fig.~\ref{fig:F_L} in
Sec.\ref{sec:F2p}.
Although significant resonant strength is observed in $F_L$
(or $R$), evidence of duality is nonetheless observed in this
structure function, along with $F_2$ and $F_1$.

In addition to the proton data, Jefferson Lab experiment E02-109
measured the LT separated $F_2$ and $F_L$ structure functions of
the deuteron, in the same $W^2$ and $Q^2$ ranges, and with the same
high precision as E94-110 did for the proton.
This will allow quantitative studies of duality in both the
longitudinal and transverse channels for the deuteron.
If duality holds well for both the proton and neutron separately,
it will hold to even better accuracy for the deuteron since the
Fermi motion effects intrinsically perform some of the averaging
over the resonances.
However, if duality does not hold for the LT separated neutron
structure functions, this should be observable in the deuteron data,
and will thus provide a critical test for models of duality.

In addition to the resonance region, measurements of the inclusive
longitudinal proton and deuteron structure have also been performed
at lower $x$ (higher $W^2$) and lower $Q^2$.
While the longitudinal strength is significant at $Q^2$ of several
GeV$^2$, the proton $F_L$ structure function is constrained by
current conservation to behave, for fixed $W$, as $F_L \sim Q^4$
for $Q^2 \to 0$.
However, even with the new Jefferson Lab data, which extend down
to $Q^2 = 0.15$~GeV$^2$ \cite{TvaskisLT}, the $Q^2$ at which this
behavior sets in has not yet been observed.

Another interesting test provided by the E99-118 data is whether the
relative longitudinal contribution to the cross section embodied in
$R$ is different in the deuteron and proton at these low $Q^2$ values.
While the higher $Q^2$ data from SLAC and NMC exhibit no significant
difference in the deuteron and proton $R$, the Jefferson Lab results
shown in Fig.~\ref{fig:rd-rp} (right) suggest a possible suppression
of $R$ in the deuteron relative to the proton for $Q^2 < 1$~GeV$^2$.
Although this suppression is consistent with the two lowest $Q^2$
data points from SLAC, the uncertainties are dominated by systematic
errors and the combined significance of the effect is still less than
2~$\sigma$.
Conclusive experimental evidence for the possible suppression of $R$
in deuterium at low $Q^2$ will likely be provided when the results
from additional data from E00-002 are finalized in very near future.

\section{Semi-inclusive deep inelastic scattering}
\label{sec:other}

In addition to the traditional $F_2$ and $F_L$ structure function
observables, a number of other processes have been studied at Jefferson
Lab over the past decade, with potentially important consequences for
our understanding of the workings of QCD at low energy.
In this section we focus on semi-inclusive pion electroproduction.

An analysis of the semi-inclusive process $e\, N \to e\, h\, X$,
where a hadron $h$ is detected in the final state in coincidence with
the scattered electron, has recently been made using data from Hall~C
in the resonance--scaling transition region \cite{Navasardyan,Mkrtchyan}.
One of the main motivations for studying semi-inclusive meson production
is the promise of flavor separation via tagging of specific mesons in
the final state.  In the valence quark region a produced $\pi^+$
($\pi^-$) meson, for example, primarily results from scattering off
a $u$ ($d$) quark in the proton.

The semi-inclusive cross section at LO in $\alpha_s$ is given by
a simple product of quark distribution and quark $\to$ hadron
fragmentation functions,
\begin{eqnarray}
{ d\sigma \over dx dz }
&\sim&\
\sum_q e_q^2\, q(x)\, D_q^h(z)\
\equiv\ {\cal N}_N^h(x,z)\, .
\label{eq:semi-parton}
\end{eqnarray}
Here the fragmentation function $D_q^h(z)$ gives the probability
for a quark $q$ to fragment to a hadron $h$ with a fraction
$z = p_h \cdot p / q \cdot p = E_h / \nu$ of the quark's
(or virtual photon's) laboratory frame energy.
Although at LO the scattering and particle production mechanisms are
independent, higher order pQCD corrections give rise to non-factorizable
terms, which involve convolutions of the PDFs and fragmentation functions
with hard coefficient functions \cite{AEMP}.

For hadrons produced collinearly with the virtual photon, the invariant 
mass $W^\prime$ of the undetected hadronic system $X$ at large $Q^2$
can be written \cite{ACW}
	$W^{\prime 2} \approx M^2 + Q^2(1-z)(1-x)/x$,
where the hadron mass is neglected with respect to $Q^2$.  In the
elastic limit, $z \to 1$, the hadron carries all of the photon's
energy (with $W^\prime \to M$), so $z$ is also referred to as the
``elasticity''.

While formally the LO factorized expression for the cross section
(\ref{eq:semi-parton}) may be valid at large $Q^2$, at finite $Q^2$
there are important corrections arising from the finite masses of
the target and produced hadron.  One can show, however, that the
LO factorization holds even at finite $Q^2$, provided the parton
distribution and fragmentation functions are expressed in terms
of generalized scaling variables \cite{AHM},
$q(x) D_q^h(z) \to q(\xi_h) D_q^h(\zeta_h)$,
where 
$\zeta_h = (z_h \xi / 2 x)
	   (1 + \sqrt{1 - 4 x^2 M^2 m_{h\perp}^2/z^2 Q^4})$
and $\xi_h = \xi (1 + m_h^2/\zeta_h Q^2)$,
with $m_{h\perp}^2 = m_h^2 + p_{h\perp}^2$.
Not surprisingly, these effects become large at large $x$ and $z$ when
$Q^2$ is small; however, for heavier produced hadrons such as kaons or
protons, significant effects can also arise at small values of $z$
\cite{AHM}.

The validity of the factorized hypothesis in Eq.~(\ref{eq:semi-parton})
relies on the existence of a sufficiently large gap in rapidity
$\eta = \ln\left[ (E_h-p_h^z)/(E_h+p_h^z) \right] / 2$ to allow a
clean separation of the current fragmentation region (hadrons produced
from the struck quark) from the target fragmentation region (hadrons
produced from the spectator quark system).  At high energies a gap
of $\Delta\eta \approx 2$, which is typically required for a clean
separation \cite{Berger}, can be achieved over a large range of $z$; at
low energies, however, this can only be reached at larger values of $z$.
On the other hand, at fixed $x$ and $Q^2$ the large-$z$ region
corresponds to resonance dominance of the undetected hadronic system
$X$ (corresponding to small $W^\prime$), so that the factorized
description in terms of partonic distributions must eventually break
down.  It is vital therefore to establish empirically the limits beyond
which the simple $x$ and $z$ factorization of Eq.~(\ref{eq:semi-parton})
is no longer valid.

It is intriguing in particular to observe whether $W^\prime$ can
play a role analogous to $W$ for duality in inclusive scattering,
when the undetected hadronic system $X$ is dominated by resonances
$W^\prime \ltorder 2$~GeV.  In terms of hadronic variables the
fragmentation process can be described through the excitation of
nucleon resonances, $N^*$, and their subsequent decays into pions
(or other mesons) and lower-lying resonances, $N'^*$.  The hadronic
description must be rather elaborate, however, as the production of
fast outgoing pions in the current fragmentation region at high energy
requires nontrivial cancellations of the angular distributions from
various decay channels \cite{IJMV,CI,WMsidis},
\begin{eqnarray}
{\cal N}_N^h(x,z)
&=& 
\sum_{N'^*}
\left|
\sum_{N^*} F_{\gamma^* N \to N^*}(Q^2,W^2)\
	   {\cal D}_{N^* \to N'^* h}(W^2,W'^2)\
\right|^2
\label{eq:dual_sidis}
\end{eqnarray}
where $F_{\gamma^* N \to N^*}$ is the $N \to N^*$ transition form factor,
which depends on the masses of the virtual photon and excited nucleon 
($W = M_{N^*}$), and ${\cal D}_{N^* \to N'^* h}$ is a function 
representing the decay $N^* \to N'^* h$.

A dedicated experiment (E00-018) to study duality in $\pi^\pm$
electroproduction was performed in Hall~C \cite{Navasardyan,Mkrtchyan},
in which a 5.5~GeV electron beam was scattered from proton and deuteron
targets at $Q^2$ between 1.8 and 6.0 GeV$^2$, for $0.3 \le x \le 0.55$,
with $z$ in the range $0.35 - 1$.  From the deuterium data the ratio of
unfavored to favored fragmentation functions $D^-/D^+$ was constructed,
where $D^+$ corresponds to a pion containing the struck quark
({\em e.g.}, $\pi^+$ from a struck $u$ or $\bar d$ quark), while $D^-$
describes the fragmentation of a quark not contained in the
valence structure of the pion ({\em e.g.}, a $d$ quark for the $\pi^+$).
Since at moderate $x$ the dependence on PDFs cancels, the fragmentation
function ratio is approximately given by
$D^-/D^+ = (4 - {\cal N}_d^{\pi^+}/{\cal N}_d^{\pi^-})/
           (4 {\cal N}_d^{\pi^+}/{\cal N_d}^{\pi^-} - 1)$,
where ${\cal N}_d^\pi$ is the yield of produced pions in
Eq.~(\ref{eq:dual_sidis}).

\begin{figure}[htb]
\begin{center}
\includegraphics[height=4.5cm]{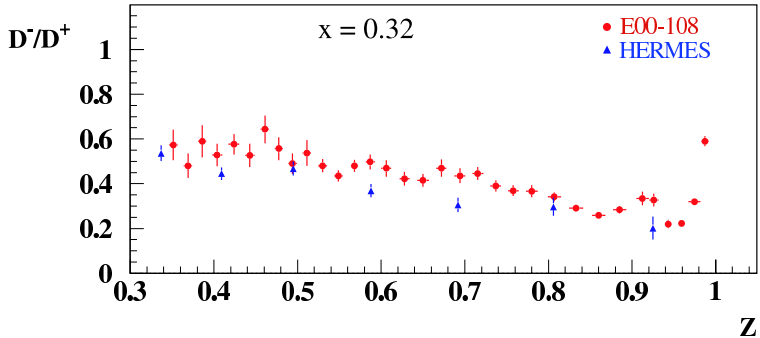}
\caption{\label{fig:dminusdplus}
	The ratio of unfavored to favored fragmentation functions
	$D^-/D^+$ as a function of $z$ extracted from deuterium data,
	for $x = 0.32$ \cite{Navasardyan}.}
\end{center}
\end{figure}

The Jefferson Lab data for $D^-/D^+$ from E00-108 are shown in
Fig.~\ref{fig:dminusdplus} as a function of $z$ at fixed $x = 0.32$
and $Q^2 = 2.3$~GeV$^2$ \cite{Navasardyan}, and compared with earlier
HERMES data at higher energies \cite{HERMES}.  Despite the different
energies, there is good overall agreement between the two measurements,
even though the Jefferson Lab data sit slightly higher.
Furthermore, the $D^-/D^+$ ratio extracted from the Jefferson Lab data
shows a smooth dependence on $z$, which is quite remarkable given that
the data cover the full resonance region,
$0.88 < W^{\prime 2} < 4.2$~GeV$^2$.
This strongly suggests a suppression or cancellation of the resonance 
excitations in the $\pi^+/\pi^-$ cross section ratio, and hence in the 
fragmentation function ratio.

Similar cancellations between resonances naturally arises in quark
models, such as those discussed by Close {\it et al.} \cite{CI,CM09}
for the $\gamma N \to \pi^\pm N'^*$ reaction.
The pattern of constructive and destructive interference, which was
a crucial feature of the appearance of duality in inclusive structure
functions, is also repeated in the semi-inclusive case when one sums
over the states $N'^*$.
Moreover, the smooth behavior of the fragmentation function ratio
$D^-/D^+$ in Fig.~\ref{fig:dminusdplus} can be qualitatively understood
from the relative weights of the matrix elements, which are always 4
times larger than for $\pi^-$ production.  In this case the resonance
contributions to this ratio cancel exactly, leaving behind only the
smooth background as would be expected at high energies.  This may
account for the striking lack of resonance structure in the resonance
region fragmentation functions in Fig.~\ref{fig:dminusdplus}.

\section{Outlook}
\label{sec:outlook}

\subsection{Impact of Jefferson Lab data}
\label{ssec:impact}

The first decade of unpolarized structure function measurements at
Jefferson Lab has had significant impact on our understanding of
nucleon structure, both for leading twist parton distributions and for
the resonance--scaling transition and related studies of quark-hadron
duality.  With most of the data concentrated in the low-$W$ region in
the $Q^2 \sim$~few GeV$^2$ range, the greatest influence on the global
data base has naturally been at large $x$.

\begin{figure}[ht]
\begin{center}
\vspace*{0.5cm}
\includegraphics[scale=0.55]{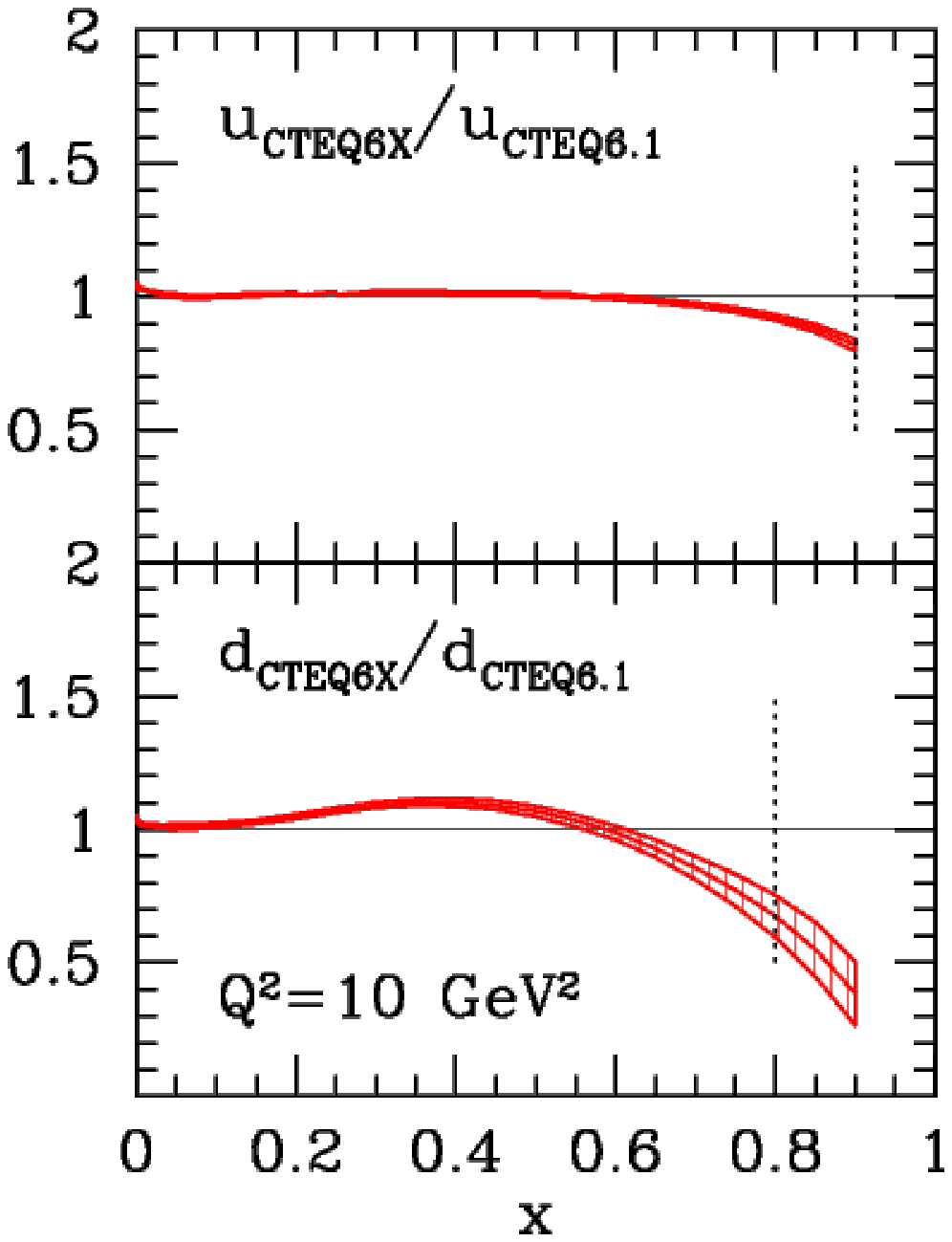}\hspace{-0cm}
\includegraphics[scale=0.5]{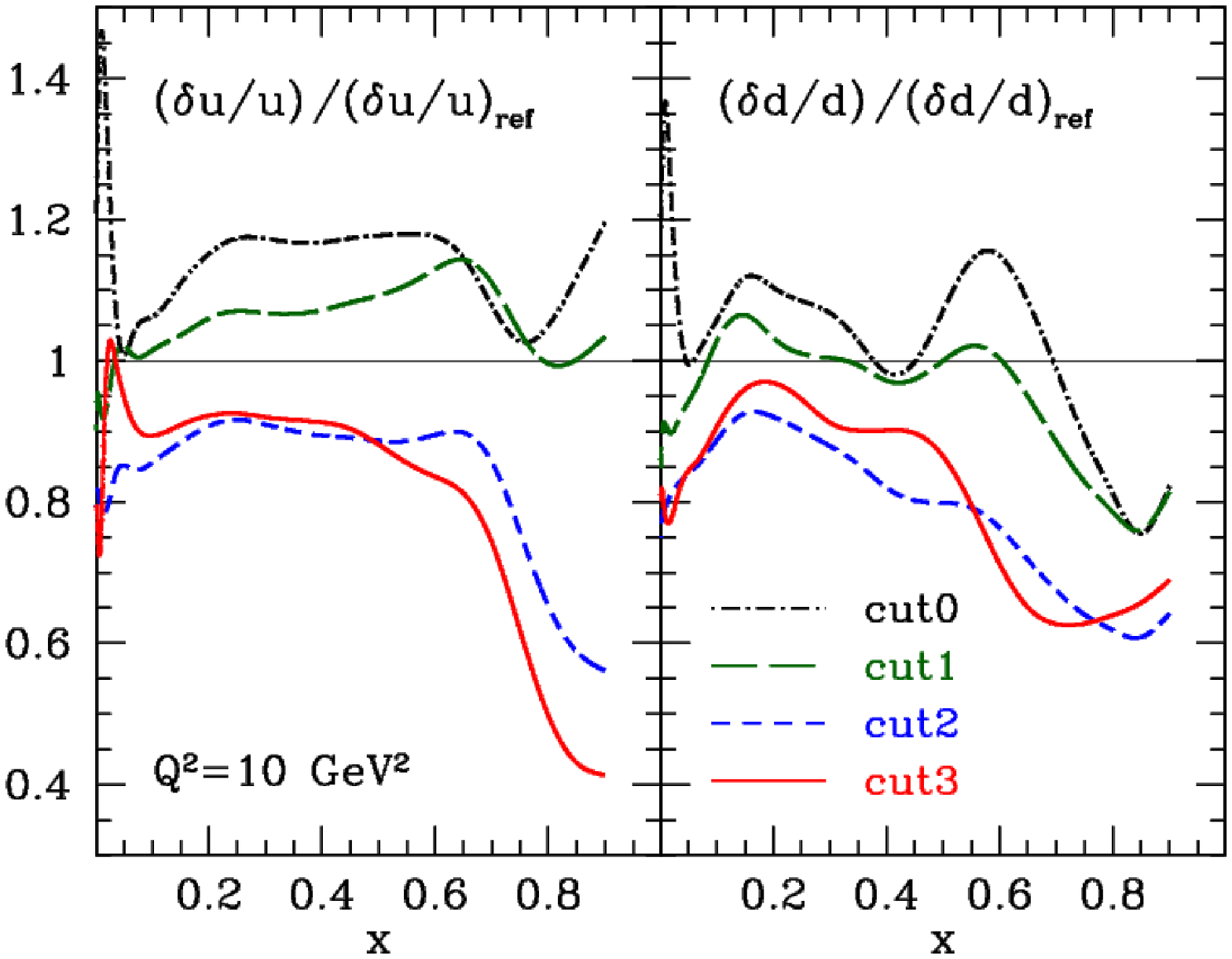}
\end{center}
\caption{{\it (Left)}
	CTEQ6X fit for $u$ and $d$ quark PDFs, normalized to the earlier
	CTEQ6.1 fit \cite{CTEQ6.1}.  The vertical lines show the
        approximate values of $x$ above which PDFs are not directly
        constrained by data.  The error bands correspond to
	$\Delta\chi=1$.
        {\it (Right)}
	Relative errors on $u$ and $d$ quark PDFs, normalized to
        the relative errors in the reference fit.}
\label{fig:cteq6x}
\end{figure}

Recently a new global PDF analysis was performed \cite{CTEQ6X}, exploring
the possibility of reducing the uncertainties at large $x$ by relaxing
the constraints on the kinematics over which data are included in the fit.
The data sets combined proton and deuteron DIS structure functions from
Jefferson Lab, SLAC and CERN (NMC) with new $ep$ collider data from HERA,
as well as new Drell-Yan, $W$~asymmetry and jet cross sections from $pp$
and $pd$ collisions at Fermilab.  The new fit (referred to as ``CTEQ6X'')
allowed for a significant increase in the large-$x$ data set ({\em e.g.},
a factor of two more DIS data points) by incorporating data for
$W^2 > 3$~GeV$^2$ and $Q^2 > 1.69$~GeV$^2$, lower than in the standard
global fits \cite{MSTW,CTEQ6.1} which typically use $W^2 > 12.25$~GeV$^2$
and $Q^2 > 4$~GeV$^2$.  The new analysis also systematically studied
the effects of target mass and higher twist contributions, and realistic
nuclear corrections for deuterium data.

Results from the CTEQ6X fit are shown in Fig.~\ref{fig:cteq6x}~(left)
for the $u$ and $d$ quark PDFs, normalized to the earlier CTEQ6.1 fit,
which had no nuclear or subleading $1/Q^2$ corrections applied.
The biggest change is the $\sim 30$--40\% suppression of the $d$ quark
at $x \sim 0.8$, which is found to be stable with respect to variations
in the $W^2$ and $Q^2$ cuts, provided both TMC and higher twist
corrections are included.  The effect of the expanded data base is more
dramatically illustrated in Fig.~\ref{fig:cteq6x}~(right), which shows
the relative $u$ and $d$ PDF errors for a range of $W$ and $Q^2$ cuts
(``cut0'' being the standard cut, ``cut3'' the more stringent CTEQ6X cut,
and intermediate cuts ``cut1'' and ``cut2''), normalized to those of a
reference fit with ``cut0'' and no nuclear or subleading corrections.
The result is a reduction of the errors by up to 40--60\% at
$x \gtorder 0.7$, which will have a profound impact on applications
of PDFs in high energy processes, such as those as the LHC, as well
as in constraining low-energy models of quark distributions.

\subsection{Future prospects}
\label{ssec:future}

Uncertainties in PDFs will be further reduced with the availability of
data at even larger $x$ and $Q^2$ from Jefferson Lab after its 12~GeV
energy upgrade, which will determine the $d$ quark distribution up to
$x \sim 0.8$ with minimal theoretical uncertainties associated with
nuclear corrections.
Several of the planned experiments include BoNuS12 \cite{BONUS12},
which will extend to larger $x$ the earlier measurements of
low-momentum, backward protons in semi-inclusive scattering from
deuterium (Sec.~\ref{ssec:F2n_bonus});
the MARATHON experiment \cite{MARATHON}, which plans to extract
$F_2^n/F_2^p$ from the ratio of $^3$He to $^3$H structure functions,
in which the nuclear corrections cancel to within $\sim 1\%$
\cite{Afnan,Salme};
and the program of parity-violating DIS measurements on hydrogen
\cite{SOLID}, which will be sensitive to a new combination of $d/u$
in the proton, free of nuclear corrections.

In the resonance region, experiment E12-10-002 \cite{E12-10-002}
will extend proton and deuteron structure function measurements up
to $Q^2 \sim 17$~GeV$^2$ and enable tests of quark-hadron duality
over a much larger kinematic range, ultimately providing stronger
constraints on large-$x$ PDFs.
And finally, a new avenue for exploring nucleon structure at 12~GeV
will be opened up with semi-inclusive meson production experiments
\cite{E12-06-104}, which will test the factorization of scattering
and fragmentation subprocesses needed for a partonic interpretation
of semi-inclusive cross sections.
A successful program of semi-inclusive measurements tagging specific
mesons in the final state would allow unprecedented access to the flavor
dependence of PDFs in previously unexplored regions of kinematics.

\ack
We thank R.~Ent and C.~E.~Keppel for their contributions to this review
in the early stages of its development.
This work was supported by the U.S. Department of Energy under Contract
No. DE-FG02-03ER41231, and DOE contract No. DE-AC05-06OR23177, under
which Jefferson Science Associates, LLC operates Jefferson Lab.


\end{document}